\documentclass[conference]{IEEEtran}
\usepackage{cite}
\usepackage{amsmath,amssymb,amsfonts}
\usepackage{algorithmic}
\usepackage{graphicx}
\usepackage{textcomp}
\usepackage{xcolor}
\usepackage{fancyhdr}
\usepackage[hyphens]{url}
\usepackage{verbatim}
\usepackage{tikz}
\usepackage{hyperref}
\usepackage{enumitem,kantlipsum}
\usetikzlibrary{arrows}\usepackage{booktabs}   
\usepackage{subcaption} 
\usepackage{xspace}
\newcommand{\fw}{TransForm\xspace} 
\newcommand{\axiom}{invlpg\xspace} 
\usepackage{hyphenat}
\usepackage{xcolor, colortbl}
\usepackage{color,soul}
\usepackage{pgf-pie}
\usepackage{pgfplots}
\mathchardef\mhyphen="2D
\pgfplotsset{compat=1.14}
\def\BibTeX{{\rm B\kern-.05em{\sc i\kern-.025em b}\kern-.08em
    T\kern-.1667em\lower.7ex\hbox{E}\kern-.125emX}}

\title{\fw: Formally Specifying Transistency Models and Synthesizing Enhanced Litmus Tests} 
\author{\IEEEauthorblockN{Naorin Hossain\IEEEauthorrefmark{1}}
\IEEEauthorblockA{\textit{Princeton University}}
\and
\IEEEauthorblockN{Caroline Trippel\IEEEauthorrefmark{1}}
\IEEEauthorblockA{\textit{Stanford University}}
\and
\IEEEauthorblockN{Margaret Martonosi}
\IEEEauthorblockA{\textit{Princeton University}}
}

\begin{document}
\maketitle
\pagestyle{plain}


\begin{abstract}
Memory consistency models (MCMs) specify the legal ordering and visibility of shared memory accesses in a parallel program.
Traditionally, instruction set architecture (ISA) MCMs assume that relevant \emph{program-visible} memory ordering behaviors only result from shared memory interactions that take place between \emph{user-level program instructions.}
This assumption fails to account for virtual memory (VM) implementations that may result in additional shared memory interactions between user-level program instructions and both 1) system-level operations (e.g., address remappings and translation lookaside buffer invalidations initiated by system calls) and 2) hardware-level operations (e.g., hardware page table walks and dirty bit updates) during a user-level program's execution.
These additional shared memory interactions can impact the observable memory ordering behaviors of user-level programs.
Thus, \textit{memory transistency models} (MTMs) have been coined as a superset of MCMs to additionally articulate VM-aware consistency rules. 
However, no prior work has enabled formal MTM specifications, nor methods to support their automated analysis. 

To fill the above gap, this paper presents the
\fw framework.
First, \fw features an axiomatic vocabulary for formally specifying MTMs. Second, \fw includes a synthesis engine to support the automated generation of litmus tests enhanced with MTM features (i.e.,  \textit{enhanced litmus tests}, or ELTs) when supplied with a \fw MTM specification.
As a case study, we formally define an estimated MTM for Intel x86 processors, called ${\tt x86t\_elt}$, that is based on observations made by an ELT-based evaluation of an Intel x86 MTM implementation from prior work and available public documentation~\cite{coatcheck, intel:x86}.
Given ${\tt x86t\_elt}$ and a synthesis bound (on program size) as input, \fw's synthesis engine successfully produces a complete set of ELTs (for synthesis bounds that complete within one week) including relevant \emph{hand-curated} ELTs from prior work, plus over 100 more.
\end{abstract}


\section{Introduction}
\label{sec:intro}
\newcommand\blfootnote[1]{%
  \begingroup
  \renewcommand\thefootnote{}\footnote{#1}%
  \addtocounter{footnote}{-1}%
  \endgroup
}

\begin{figure}
    \centering
    \includegraphics[width=.53\linewidth]{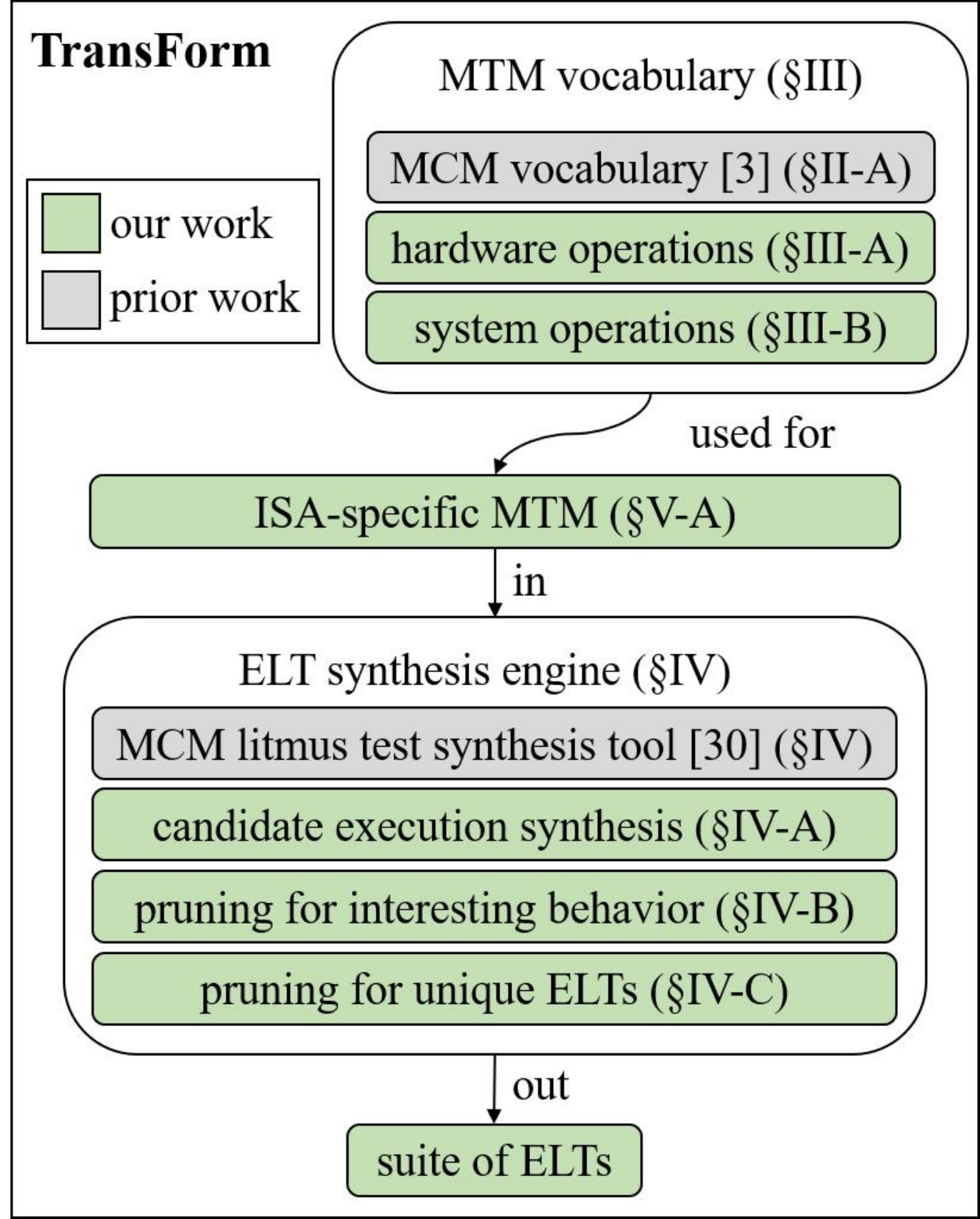}
    \caption{\fw features 1) an axiomatic vocabulary for specifying MTMs and 2) a synthesis engine for generating ELTs from MTM specifications.}
    \label{fig:block}
\end{figure}

Programmers and system designers rely on interface specifications to coordinate software's correct execution on hardware systems.\blfootnote{\IEEEauthorrefmark{1}The first two authors contributed equally to this paper.}
For example, instruction set architectures (ISAs) feature \textit{memory consistency models} (MCMs) which specify the legal orderings and visibility of shared memory accesses in any parallel program running on an implementation of the ISA.
Defining behavior as fundamental as what value can be returned when software loads from memory, both under-specified and incorrectly implemented ISA MCMs have resulted in a range of bugs in
real-world programs~\cite{ARMHazard, amd:documents, tricheck, guanciale:oakland16}, and significant effort is devoted 
to specifying them and verifying their correct implementation~\cite{pipecheck,ccicheck,coatcheck,tricheck,rtlcheck,pipeproof}.
However, this paper addresses the observation that traditional ISA MCMs
fail to capture relevant
shared memory interactions and therefore relevant memory ordering behaviors~\cite{coatcheck}.


{\bf Transistency:} MCMs assume that program-visible memory ordering behaviors only result from shared memory interactions that take place between user-level program instructions. Thus, MCMs abstract away processor virtual memory (VM) implementations that may result in additional shared memory interactions between user-level program instructions and both 1) system-level operations (e.g., address remappings and translation lookaside buffer, or TLB, invalidations initiated by system calls)
and 2) hardware-level operations (e.g., hardware page table walks and dirty bit updates) during a user-level program's execution~\cite{romanescu:vamc,romanescu:at}.
Along with involving non-user-facing instructions,
these additional shared memory interactions 
take place via shared memory state that is typically
outside the purview of MCMs. This \textit{transistency state} includes 1) page table entries (PTEs) which store virtual-to-physical address (VA-to-PA) mappings that are modifiable by the operating system (OS) and PTE status bits (e.g., dirty bits) that are modifiable directly by hardware,
and 2) TLB entries which cache VA-to-PA mappings on each core and can be evicted by the OS via inter-processor interrupts (IPIs) or loaded by hardware.
VM-specific systems behaviors can negatively impact concurrent program executions~\cite{amdphenom,amd:documents,amdtlb3,intel:xeon}, so ISA-level event ordering specifications like MCMs should include them. Failure to incorporate VM-aware features into ISA MCM specifications and subsequent hardware MCM verification implicitly makes the erroneous assumption
that underlying VM implementations will not negatively impact program correctness. 

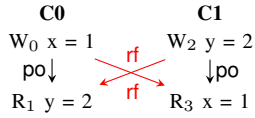
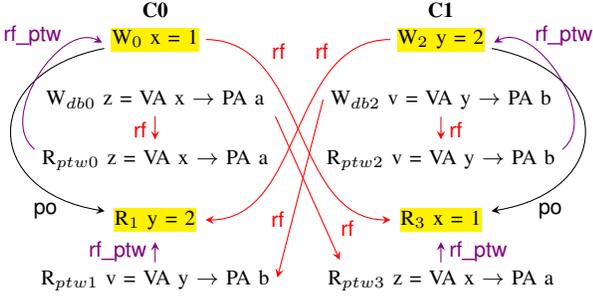
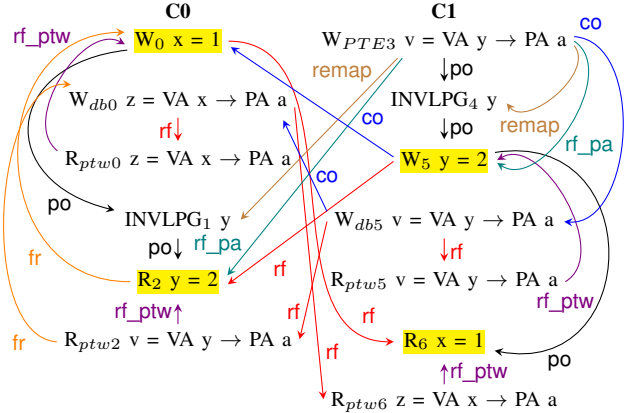
\begin{figure}[t]
\begin{minipage}[b]{\linewidth}
    \begin{subfigure}[b]{\linewidth}
    \centering

        \begin{tikzpicture}[->,>=stealth,shorten >=1pt,auto,node distance=.8cm, node/.style={rectangle,draw=none,fill=none,minimum size=1mm}]
            \footnotesize
            \node[node] (C0) {\textbf{C0}};
            \node[node] (C1) [right of=C0, xshift=1.3cm] {\textbf{C1}};
            \node[node] (i0) [below of=C0, yshift=.4cm] {W$_{0}$ x = 1};
            \node[node] (i1) [below of=i0] {R$_{1}$ y = 2};
            \node[node] (i2) [below of=C1, yshift=.4cm] {W$_{2}$ y = 2};
            \node[node] (i3) [below of=i2] {R$_{3}$ x = 1};
        
            \path[every node/.style={font=\sffamily\footnotesize, fill=none,inner sep=2pt}]
                (i0) edge [left] node {po} (i1)
                (i2) edge [right] node {po} (i3)
                (i0) edge [red] node[xshift=-4pt, yshift=2pt] {rf} (i3)
                (i2) edge [red] node[xshift=-4pt, yshift=-2pt] {rf} (i1);
        \end{tikzpicture}
        \caption{User-level representation of the {$\tt sb$} litmus test where the ${\tt Reads}$ on cores ${\tt C0}$ and ${\tt C1}$ (${\tt R_1}$, ${\tt R_3}$) return the values 2 and 1, respectively.}
        \label{fig:sb}
    \end{subfigure}
\end{minipage} \\
\begin{minipage}[b]{\linewidth}
    \begin{subfigure}[b]{\linewidth}
    \centering
	    \begin{tikzpicture}[->,>=stealth,shorten >=1pt,auto,node distance=.8cm, node/.style={rectangle,draw=none,fill=none,minimum size=1mm}]
              \footnotesize
              \node[node] (C0) {\textbf{C0}};
              \node[node] (C1) [right of=C0, xshift=3cm] {\textbf{C1}};
              \node[node] (i1) [below of=C0, yshift=.4cm] {\hl{W$_{0}$ x = 1}};
              \node[node] (i2) [below of=i1]   {W$_{db0}$ z = VA x $\rightarrow$ PA a};
              \node[node] (i3) [below of=i2]   {R$_{ptw0}$ z = VA x $\rightarrow$ PA a};
              \node[node] (i4) [below of=i3]   {\hl{R$_{1}$ y = 2}};
              \node[node] (i5) [below of=i4]   {R$_{ptw1}$ v = VA y $\rightarrow$ PA b};
              \node[node] (i7) [below of=C1, yshift=.4cm]  {\hl{W$_{2}$ y = 2}};
              \node[node] (i8) [below of=i7]   {W$_{db2}$ v = VA y $\rightarrow$ PA b};
              \node[node] (i9) [below of=i8]   {R$_{ptw2}$ v = VA y $\rightarrow$ PA b};
              \node[node] (i10) [below of=i9]   {\hl{R$_{3}$ x = 1}};
              \node[node] (i11) [below of=i10]   {R$_{ptw3}$ z = VA x $\rightarrow$ PA a};
            
              \path[every node/.style={font=\sffamily\footnotesize, fill=none,inner sep=3pt}]
                (i1) edge [bend right=80, looseness=2, very near end, left] node[yshift=-4pt] {po} (i4)
                (i7) edge [bend left=80, looseness=2, very near end, right] node[yshift=-4pt] {po} (i10)
                
                (i2) edge [red, left] node {rf} (i3)
                (i3) edge [violet, bend left=85, near end, left] node[yshift=4pt] {rf\_ptw} (i1)
                (i5) edge [violet, left] node {rf\_ptw} (i4)
                
                (i1) edge [red, near start, out=0, in=180] node {rf} (i10)
                (i7) edge [red, near start, left, above, out=180, in=0] node[xshift=-4pt] {rf} (i4)
                (i2.-10) edge [red, xshift=10pt, very near end] node {rf} (i11.170)
                (i8.180) edge [red, near end, left, above] node[xshift=-4pt] {rf} (i5.0)
                
                (i8) edge [red, right] node {rf} (i9)
                (i9) edge [violet, bend right=85, near end, right] node[yshift=4pt] {rf\_ptw} (i7)
                (i11) edge [violet, right] node {rf\_ptw} (i10);
        \end{tikzpicture}
        \caption{{$\tt sb$} mapped to an ELT where the outcome remains permitted.}
        \label{fig:sb-elt-permitted}
    \end{subfigure}
\end{minipage} \\
\begin{minipage}[b]{\linewidth}
\begin{subfigure}[b]{\linewidth}
        \centering
        \begin{tikzpicture}[->,>=stealth,shorten >=1pt,auto,node distance=.8cm, node/.style={rectangle,draw=none,fill=none,minimum size=1mm}]
          \footnotesize
          \node[node] (C0) {\textbf{C0}};
          \node[node] (C1) [right of=C0, xshift=2.74cm] {\textbf{C1}};
          \node[node] (i1) [below of=C0, yshift=.4cm] {\hl{W$_{0}$ x = 1}};
          \node[node] (i2) [below of=i1]   {W$_{db0}$ z = VA x $\rightarrow$ PA a};
          \node[node] (i3) [below of=i2]   {R$_{ptw0}$ z = VA x $\rightarrow$ PA a};
          \node[node] (i30) [below of=i3]   {INVLPG$_{1}$ y};
          \node[node] (i4) [below of=i30]   {\hl{R$_{2}$ y = 2}};
          \node[node] (i5) [below of=i4]   {R$_{ptw2}$ v = VA y $\rightarrow$ PA a};
          
          \node[node] (i60) [below of=C1, yshift=.4cm]   {W$_{PTE3}$ v = VA y $\rightarrow$ PA a};
          \node[node] (i61) [below of=i60]   {INVLPG$_{4}$ y};
          \node[node] (i7) [below of=i61]   {\hl{W$_{5}$ y = 2}};
          \node[node] (i8) [below of=i7]   {W$_{db5}$ v = VA y $\rightarrow$ PA a};
          \node[node] (i9) [below of=i8]   {R$_{ptw5}$ v = VA y $\rightarrow$ PA a};
          \node[node] (i10) [below of=i9]   {\hl{R$_{6}$ x = 1}};
          \node[node] (i11) [below of=i10]   {R$_{ptw6}$ z = VA x $\rightarrow$ PA a};
        
          \path[every node/.style={font=\sffamily\footnotesize, fill=none,inner sep=2pt}]
            (i1) edge [bend right=80, looseness=2, very near end, left] node[yshift=-4pt] {po} (i30)
            (i30) edge [left] node {po} (i4)
            (i61) edge [right] node {po} (i7)
            (i7) edge [bend left=100, looseness=2, very near end] node {po} (i10)
            
            (i2) edge [red, left] node {rf} (i3)
            (i3) edge [violet, bend left=85, near end, left] node[yshift=4pt, xshift=-6pt] {rf\_ptw} (i1)
            (i5) edge [violet, left] node {rf\_ptw} (i4)
            
            (i1) edge [red, very near end, out=0, in=180] node {rf} (i10)
            (i7.180) edge [red, near end] node {rf} (i4.-5)
            (i2.-3) edge [red, very near end] node {rf} (i11.180)
            (i8.180) edge [red, near end, left] node {rf} (i5.0)
            
            (i8) edge [red, right] node {rf} (i9)
            (i9) edge [violet, bend right=90, at start, right, below] node[yshift=-2pt] {rf\_ptw} (i7)
            (i11) edge [violet, right] node {rf\_ptw} (i10)
            
            (i60) edge [right] node {po} (i61)
            (i60.-2) edge [brown, bend left=90, very near end, below, right] node[yshift=-8pt, xshift=-8pt] {remap} (i61.0)
            (i60.200) edge [brown, very near start, left] node[yshift=2pt] {remap} (i30.0)
            (i60.202) edge [teal, very near end, left] node[yshift=2pt] {rf\_pa} (i4.5)
            (i60) edge [teal, bend left=90, right] node[yshift=-8pt, xshift=-4pt] {rf\_pa} (i7)
            (i8.175) edge [blue] node[xshift=15pt] {co} (i2.-9)
            (i60.2) edge [blue, bend left=90, right, at start] node[yshift=5pt] {co} (i8.0)
            (i7.175) edge [blue, very near start, above] node[yshift=5pt] {co} (i1.-5)
            (i4.180) edge [orange, bend left=90, looseness=1.5, near start] node {fr} (i1.170)
            (i5.180) edge [orange, bend left=90, looseness=.8, very near start] node {fr} (i2.171)
            ;
        \end{tikzpicture}
        \caption{{$\tt sb$} mapped to an ELT where the outcome is now forbidden due to VAs {$\tt x$} and {$\tt y$} aliasing to the same PA {$\tt a$}.}\label{fig:sb-elt-forbidden}
    \end{subfigure}
\end{minipage}
\caption{(\subref{fig:sb}) illustrates a sequentially consistent execution of the {$\tt sb$} litmus test using traditional MCM annotations. (\subref{fig:sb-elt-permitted})~and~(\subref{fig:sb-elt-forbidden}) show two possible mappings of {$\tt sb$} to ELTs using annotations representative of our new MTM vocabulary. User-facing instructions are highlighted in yellow.}
\label{fig:motivating}
\end{figure}

To augment MCMs with VM-aware features, prior work proposed {\em memory transistency models} (MTMs): ``the superset of [memory] consistency [models] which capture all [address] translation-aware sets of ordering rules''\cite{coatcheck}.   Likewise, where ISA MCM behaviors are typically specified and validated using small diagnostic programs called litmus tests (such as ${\tt sb}$ in Fig.~\ref{fig:sb})~\cite{intel:x86,amd:documents,alglave:herd,tsotool,Alglave:litmus,Mador-Haim:litmus,lustig:automated}, MTMs use \textit{enhanced litmus tests} (ELTs) as a mechanism for encoding and testing the effects of VM operations on parallel program execution. ELTs are small parallel programs, comprised not just of user-facing ISA-level events (i.e., ISA instructions or micro-ops), but also of system- and hardware-level events that execute on behalf of or interleaved with user-facing instructions. Figs.~\ref{fig:sb-elt-permitted}~and~\ref{fig:sb-elt-forbidden} (explained in \S\ref{sec:mcm_limits}) give examples of ELTs, which, in contrast to Fig.~\ref{fig:sb}'s standard MCM litmus test, include system- and hardware-level operations that access program-visible transistency state.
Unfortunately, no prior work has formally defined MTMs, an essential step for enabling automated ELT generation and thus ELT-based validation and verification of MTM implementations. Furthermore, the ELTs of prior work were largely hand-generated\footnote{Prior work automates the insertion of ghost instructions (\S\ref{sec:ghost}) into hand-generated ELTs based on user-defined rules.}, and therefore incomplete.

This paper presents the \textit{\fw framework} (short for \textit{transistency formalized}) to support the development of formally specified MTMs, and to automate the synthesis of ELTs to validate them.
\fw (Fig.~\ref{fig:block}) consists of 1) an \emph{axiomatic vocabulary} for formally defining arbitrary MTMs, and 2) a \emph{synthesis engine} for automatically generating ELTs from ISA-level MTM specifications defined using the \fw vocabulary. \fw's axiomatic vocabulary extends beyond the standard axiomatic MCM vocabulary~\cite{alglave:herd,batty:overhauling,mador-haim:axiomatic,armpowertutorial}. It provides new constructs for modeling MTM-specific features such as particular VM-relevant shared memory state and additional MTM-relevant system- and hardware-level operations (i.e., \emph{transistency operations}) that may interact with user-facing program instructions via this additional state. 

\fw's synthesis engine provides a mechanism to automate ELT synthesis from axiomatic MTM specifications. Together, formal MTMs and automated ELT synthesis support the specification and subsequent verification and validation of complex hardware-software event ordering scenarios. For example, a bug in AMD Athlon\texttrademark{} 64 and Opteron\texttrademark{} processors caused ${\tt INVLPG}$ instructions (the x86 instruction for evicting a TLB entry, which is described further in \S\ref{sec:invlpg}) to fail to invalidate the designated TLB entries~\cite{amdtlb3}. Such a bug, which could be detected by \fw-synthesized ELTs, can result in the use of a stale address mapping.

As a case study, we define ${\tt x86t\_elt}$, an estimated\footnote{``Estimated'' conveys that {$\tt x86t\_elt$} is designed to comply with a suite of hand-generated ELTs and available public documentation~\cite{coatcheck, intel:x86} which might be ambiguous, as pointed out in prior MCM work~\cite{Bornholt:memsynth}.} MTM for Intel x86 processors based on observations made by an ELT-based evaluation of an Intel x86 MTM implementation from prior work and available public documentation~\cite{intel:x86,coatcheck}.
We supply {$\tt x86t\_elt$} as input to \fw's synthesis engine to generate an ELT suite for evaluating the model's efficacy. 

We summarize our contributions as follows:

\noindent
\textbf{\fw:} We present the \fw framework for formally defining MTMs and synthesizing ELTs from MTM specifications to support MTM verification and validation. 

\noindent
\textbf{Axiomatic MTM vocabulary:} 
To the best of our knowledge, there have been no prior efforts to formally specify MTMs.
TransFrom augments MCMs with 1) transistency operations and 2) relations for articulating shared memory interactions between transistency operations and user-facing instructions.

\noindent
\textbf{ELT synthesis:} \fw automates ELT generation from formal MTM specifications written in \fw's vocabulary. The synthesized ELTs can be used to automate MTM verification and  validation and ultimately inform system designers about the software-visible effects of VM implementations.

\noindent
\textbf{x86 case study:} 
We define an estimated MTM for Intel x86 processors, called {$\tt x86t\_elt$}, based on publicly-available documentation and prior ELT-based evaluation~\cite{intel:x86, coatcheck}. Given {$\tt x86t\_elt$}, \fw synthesizes all ELTs satisfying defined criteria for minimality and stressing behaviors of the model. \fw's synthesis engine successfully produces a complete set of ELTs (for synthesis bounds that complete within one week) including all relevant \emph{hand-curated} ELTs from prior work~\cite{coatcheck} plus over 100 more. This evaluation constitutes the first automatically-synthesized and largest set of ELTs that can be used for validating Intel x86 MTM implementations. 

\section{Background {\&} Motivation}

Since MCMs are a fundamental component for reasoning about parallel program correctness, there has been significant prior work on formally specifying them~\cite{manson:java, boehm:cppconcurrency, petri:cooking, Batty:mathematizingc++, batty:overhauling, nienhuis:c11operational, Wickerson2015, nagarajan2020primer, owens:better, alglave:herd, pulte:armv8, nvidia:ptx, RISCV:rvtso:rvwmo}. Much of this work uses axiomatic-style (i.e. declarative) specifications, which describe the legal executions of a program with the help of logical axioms. These axioms encode the conditions that must hold true during any execution under the defined MCM.
\S\ref{sec:mcm-vocab} gives an overview of axiomatic ISA MCM specifications and a standard vocabulary for defining them.  \S\ref{sec:mcm_limits} highlights the limitations of this vocabulary for capturing MTM-relevant program behaviors. \S\ref{sec:towards} extends the vocabulary in \S\ref{sec:mcm-vocab} to incorporate transistency operations.

\subsection{Axiomatic Vocabulary for Specifying Memory Models}
\label{sec:mcm-vocab}

Two primary sets, referred to here as ${\tt Event}$ and ${\tt Location}$, can serve as the basis for defining ISA MCMs. ${\tt Event}$ is the set of all micro-ops (typically memory and synchronization operations) in a given program execution. ${\tt Location}$ is the set of all memory locations. Referring to the \emph{store buffering} (${\tt sb}$) litmus test in Fig.~\ref{fig:sb}, ${\tt Event = \{W_{0}, R_{1}, W_{2}, R_{3}\}}$ and ${\tt Location = \{x, y\}}$. ${\tt MemoryEvent}$ is a subset of ${\tt Event}$, containing only micro-ops that access memory (e.g., via reading or writing it). In Fig.~\ref{fig:sb}, all elements of ${\tt Event}$ are also elements of ${\tt MemoryEvent}$. ${\tt MemoryEvent}$ can be further divided into ${\tt Read}$ and ${\tt Write}$ subsets that contain micro-ops that read and write memory, respectively. Each ${\tt MemoryEvent}$ element is related to exactly one ${\tt Location}$ element by the ${\tt address}$ relation. In this paper's litmus test examples, the notation \textless{}${\tt mem\_op}$\textgreater{} \textless{}${\tt addr}$\textgreater{} = \textless{}${\tt data}$\textgreater{} indicates that 
${\tt mem\_op}$ is related to
${\tt addr}$ by the ${\tt address}$ relation. In Fig.~\ref{fig:sb}, ${\tt address = \{(W_{0}, x), (R_{1}, y), (W_{2}, y), (R_{3}, x)\}}$.

Relations can be denoted as labels (e.g., ${\tt address}$ encodes a labeling of ${\tt MemoryEvents}$ with ${\tt Locations}$) or directed edges in program execution graphs. Directed edges indicate sequencing relationships between ${\tt Events}$. We describe some baseline MCM ``edge relations'' here noting that others may be derived from this baseline set as needed. ${\tt Events}$ that are sequenced in \textit{program order} are related by the ${\tt po}$ relation. In Fig.~\ref{fig:sb}, earlier instructions are related to subsequent same-thread instructions by ${\tt po}$, denoted by directed ${\tt po}$ edges. Additionally, ${\tt MemoryEvents}$ that access \textit{the same} memory location can be related by the \textit{reads-from} (${\tt rf}$), \textit{coherence-order} (${\tt co}$), or \textit{from-reads} (${\tt fr}$) relations. ${\tt rf}$ relates ${\tt Writes}$ to ${\tt Reads}$ that they source;
${\tt co}$ relates ${\tt Writes}$ to other ${\tt Writes}$ that come later in coherence order (i.e., ${\tt co}$ is a total order on same-address ${\tt Writes}$); and ${\tt fr}$ relates {$\tt Reads$} to ${\tt Writes}$ that are ${\tt co}$-successors of the ${\tt Write}$ they read from.
We refer to the union of ${\tt rf}$, ${\tt co}$, and ${\tt fr}$ as \textit{communication} (${\tt com}$) relations.

A given set of {$\tt Event$} and {$\tt Location$} elements along with a set of {$\tt address$} and {$\tt po$} relations defines a \textit{program}. Adding {$\tt com$} relations (which distinguish different executions of the same program)
defines a \textit{candidate execution}---i.e., a possible dynamic sequencing of program memory references and other MCM-relevant operations (e.g., synchronization operations like fences/barriers). A litmus test, as in Fig.~\ref{fig:sb}, depicts a candidate execution.
In essence, ${\tt com}$ relations encode final \textit{outcomes} of litmus test programs, where an outcome consists of the values returned by program ${\tt Reads}$ and the final state of memory. \fw represents all stored values (and thus outcomes of candidate executions) symbolically.
However, the examples in this paper feature concrete values for pedagogy.

An MCM specification defines a \textit{consistency predicate} that renders candidate executions \textit{consistent} or \textit{inconsistent} with respect to the specification. For example, the total store order (TSO) MCM used by Intel x86~\cite{intel:x86} processors, known as ${\tt x86\mhyphen{}TSO}$, is defined by a consistency predicate that is composed of the conjunction of three axioms (i.e., predicates that must evaluate to ${\tt True}$): {$\tt sc\_per\_loc$}, {$\tt rmw\_atomicity$}, and {$\tt causality$}~\cite{alglave:herd}. These are defined as follows:
\begin{enumerate}
    \item ${\tt sc\_per\_loc}$: The set $\{{\tt rf} + {\tt co} + {\tt fr} + {\tt po\_loc}\}$ of edges, where $+$ indicates disjunction, is acyclic. ${\tt po\_loc}$ is the subset of ${\tt ^{\wedge}po}$ that relates same-address ${\tt MemoryEvents}$, where ${\tt ^{\wedge}}$ is the transitive closure operator.
    \item ${\tt rmw\_atomicity}$: There are no intervening same-address ${\tt Writes}$ between the ${\tt Read}$ and ${\tt Write}$ of a read-modify-write (RMW) operation. 
    In other words, ${\tt fr}.{\tt co}$ does not intersect with ${\tt rmw}$, where relation ${\tt rmw}$ relates the ${\tt Read}$ of an RMW to its corresponding ${\tt Write}$, and where $.$ is the join operator.
    \item ${\tt causality}$: The set $\{{\tt rfe} + {\tt co} + {\tt fr} + {\tt ppo} + {\tt fence}\}$ of edges is acyclic. 
    \textit{Preserved program order} (${\tt ppo}$)  corresponds to a subset of ${\tt ^{\wedge}po}$ where the sequencing order denoted by ${\tt ^{\wedge}po}$ must be maintained by the architecture. ${\tt fence}$ relates ${\tt Events}$ whose ordering is explicitly architecturally-enforced by the presence of fence or barrier ${\tt Events}$. \textit{Reads-from external} (${\tt rfe}$) is the subset of ${\tt rf}$ that relates ${\tt Events}$ on different threads.
\end{enumerate}

\subsection{Limitations of Current ISA Memory Models}
\label{sec:mcm_limits}

While MCMs of today's commercial hardware~\cite{amd:documents, intel:itanium, intel:x86, ibm:power, ARMv7, nvidia:ptx, ARMv8, RISCV:rvtso:rvwmo} are fundamental for precisely specifying the legal ordering and visibility of shared memory accesses in a parallel program, there are ways in which they are insufficient.
Central to our work, ISA memory and synchronization operations that are fetched, decoded, and issued as part of the user-level instruction stream are \emph{not} the only operations that may affect the outcome of a user-level program. Thus, our work encompasses typical consistency features as well as \emph{transistency features}.
In particular, our work on MTMs additionally captures shared memory interactions between user-level instructions and transistency operations (i.e., system-level and hardware-level operations). 
The system-level operations (i.e., \emph{support instructions}\footnote{Prior work encompasses TransForm's support operations---address remappings and TLB invalidations initiated by system calls---in coarser-grained
map-remap functions (MRFs)~\cite{romanescu:vamc}.}) we consider include  address  remappings  and  TLB  invalidations  initiated by system calls.
The hardware-level operations (i.e., \emph{ghost instructions}~\cite{coatcheck}) we consider include hardware page table walks (PT walks) and dirty bit updates.
Furthermore, our work supports modeling of MTM-specific shared memory interactions by expanding the notion of ``data'' from MCMs beyond program variables to also include
transistency-specific data (i.e., transistency state) like page table dirty bits and VA-to-PA mappings themselves.

\subsubsection{Transistency Impacts Program Executions}

Fig.~\ref{fig:motivating} motivates augmenting MCMs with transistency features. For the MCMs of essentially all commercial processors, the litmus test execution in Fig.~\ref{fig:sb} is perfectly legal. In fact, Fig.~\ref{fig:sb} features a sequentially-consistent execution~\cite{lamport:sc}. However, accounting for transistency could render the litmus test execution illegal
if, for example, virtual addresses (VAs) ${\tt x}$ and ${\tt y}$ were to map to the same physical address (PA). 

Figs.~\ref{fig:sb-elt-permitted}~and~\ref{fig:sb-elt-forbidden} represent two possible ways in which transistency features could affect the legality of the execution of Fig.~\ref{fig:sb}. The program executions in these figures that are enhanced with transistency features are ELTs~\cite{coatcheck}. Before explaining these ELTs, we state some assumptions made in the litmus tests we present.
First, as is typical for litmus tests, memory locations in ELTs are initialized at the start of the test. Thus, a ${\tt Read}$ that is not involved in an ${\tt rf}$ relation reads from the initial program state. In keeping with MCM convention, program variables are initialized to 0 at the start of the test. 
Furthermore, the ELTs we present assume the following initial mappings in PTEs stored at VAs ${\tt z}$ and ${\tt v}$: VA ${\tt z}$: VA ${\tt x}$ $\rightarrow$ PA ${\tt a}$ and VA ${\tt v}$: VA ${\tt y}$ $\rightarrow$ PA ${\tt b}$, respectively. Again, all shared memory values are represented symbolically by \fw and concretely in our examples for pedagogy. Next,  ${\tt x}$, ${\tt y}$, and ${\tt u}$ are VAs and  ${\tt a}$, ${\tt b}$, and ${\tt c}$ are PAs. Finally, per-core TLBs are initially empty.

Fig.~\ref{fig:sb-elt-permitted} features one possible result of augmenting 
Fig.~\ref{fig:sb}'s execution with ${\tt Events}$ related to transistency. 
First, each user-facing ${\tt Write}$, ${\tt W}$, invokes a ghost instruction, ${\tt W_{db}}$. Each ${\tt W_{db}}$ is a ${\tt Write}$ event that accesses a shared memory ${\tt Location}$ containing the dirty bit in the PTE that corresponds to the effective VA of the shared-memory ${\tt Write}$ that invoked it. The causal relationship between the user-facing ${\tt Writes}$ and their corresponding dirty bit ${\tt Writes}$ is denoted by matching numerical subscripts (explained further in \S\ref{sec:ghost}).
Next, each ${\tt MemoryEvent}$ invokes a PT walk ghost instruction to locate the VA-to-PA mapping corresponding to its effective VA. The mapping from Fig.~\ref{fig:sb} to  Fig.~\ref{fig:sb-elt-permitted} is an algorithmic translation that expands user-level instructions to include ghost instructions executing on their behalf.  The execution it represents would be legal on essentially all commercial MCMs.

Fig.~\ref{fig:sb-elt-forbidden} shows another possible augmentation of Fig.~\ref{fig:sb}.
In this case, however, the resulting ELT now represents an illegal execution on virtually all commercial processors. The illegal behavior stems from a support operation, ${\tt W_{PTE3}}$, on Core 1 (${\tt C1}$) which modifies the VA-to-PA mapping stored at VA ${\tt v}$ and results in VAs ${\tt x}$ and ${\tt y}$ aliasing the same PA ${\tt a}$. ${\tt W_{PTE}}$ operations are ${\tt Write}$ events that result from address remapping system calls made on behalf of the user-level program. Each ${\tt W_{PTE}}$ accesses a shared memory ${\tt Location}$ containing the VA-to-PA mapping to be modified. 
The address remapping in Fig.~\ref{fig:sb-elt-forbidden} results in the ELT featuring a 
coherence violation (i.e., a violation of ${\tt sc\_per\_loc}$), thus rendering it illegal under ${\tt x86\mhyphen{}TSO}$ (defined in \S\ref{sec:mcm-vocab}).

The ELTs in Fig.~\ref{fig:motivating} illustrate that a candidate execution's legality cannot necessarily be determined solely by information provided in traditional MCM litmus tests (as in Fig.~\ref{fig:sb}). 
${\tt Events}$ and relationships related to VM implementations must be taken into account (as in ELTs) since they can impact the correctness of interactions between user-level ${\tt Events}$. A given candidate execution (or ELT)
is determined to be permitted or forbidden for a given MTM by evaluating the candidate execution against the MTM’s {\em transistency
predicate}.


\subsubsection{A Need for Formal MTM Specifications}

The prior work that proposed ELTs additionally presented a transistency-aware framework, called COATCheck, for specifying and verifying microarchitectural MTM implementations~\cite{coatcheck}.
COATCheck facilitates the specification of hardware designs along with their VM-relevant OS support in a way that is amenable to analysis with formal techniques. However, COATCheck conducts verification with respect to a user-provided suite of ELT programs that have been hand-curated.
Improving microarchitectural MTM verification coverage to more thoroughly verify correct execution of corner-case behaviors requires a way to automatically and systematically generate relevant ELTs for a given MTM. Formal specification of an ISA's MTM is required to serve as the basis for automated ELT synthesis.

\emph{Our work gives MTMs a formal semantics.} \fw offers a language for formally specifying MTMs at the \textit{architectural-level} whereas COATCheck demonstrated the importance of formal MTM verification at the \textit{microarchitectural-level}. More broadly, formally specifying an ISA's MTM provides a precise interface against which tools such as COATCheck can conduct verification of hardware implementations and programs targeting those implementations, even extending to full proofs of MTM correctness in the future~\cite{pipeproof}.

\section{Towards an MTM}
\label{sec:towards}

Starting from \S\ref{sec:mcm-vocab}'s baseline vocabulary for describing MCMs axiomatically, we propose additional ${\tt Events}$ and relations that are essential for defining MTMs and synthesizing ELTs. As we detail TransForm's transistency vocabulary, we reference Fig.~\ref{fig:motivating} as a running example.


\subsection{Hardware-Level Operations: Ghost Instructions}
\label{sec:ghost}

\begin{figure}[t]
\begin{minipage}{\linewidth}
\begin{minipage}[b]{\linewidth}
    \begin{subfigure}[b]{\linewidth}
    \centering
	    \begin{tikzpicture}[->,>=stealth,shorten >=1pt,auto,node distance=.8cm, node/.style={rectangle,draw=none,fill=none,minimum size=1mm}]
          \footnotesize
          \node[node] (C0) {\textbf{C0}};
          \node[node] (i0) [below of=C0, yshift=.4cm]   {R$_{0}$ x = 0};
          \node[node] (i1) [below of=i0]   {R$_{ptw0}$ z = VA x $\rightarrow$ PA a};

          \path[every node/.style={font=\sffamily\footnotesize, fill=none,inner sep=2pt}]
            (i1) edge [violet, left] node {rf\_ptw} (i0)
            ;
        \end{tikzpicture}
        \caption{User-facing ${\tt Reads}$ may also result in additional memory references in the form of a PT walk operation (${\tt R_{ptw0}}$).}
        \label{fig:ghost_read}
    \end{subfigure}
\end{minipage}
\\
\begin{minipage}[b]{\linewidth}
    \begin{subfigure}[b]{\linewidth}
    \centering
	    \begin{tikzpicture}[->,>=stealth,shorten >=1pt,auto,node distance=.8cm, node/.style={rectangle,draw=none,fill=none,minimum size=1mm}]
          \footnotesize
          \node[node] (C0) {\textbf{C0}};
          \node[node] (i1) [below of=C0, yshift=.4cm]   {W$_{0}$ x = 1};
          \node[node] (i2) [below of=i1]   {W$_{db0}$ z = VA x $\rightarrow$ PA a};
          \node[node] (i3) [below of=i2]   {R$_{ptw0}$ z = VA x $\rightarrow$ PA a};
          
          \path[every node/.style={font=\sffamily\footnotesize, fill=none,inner sep=2pt}]
            (i3) edge [violet, bend left=90, left] node {rf\_ptw} (i1)
            ;
        \end{tikzpicture}
        \caption{Like the ${\tt Read}$ in (\subref{fig:ghost_read}), user-facing ${\tt Writes}$ may result in a PT walk (${\tt R_{ptw0}}$). Additionally, ${\tt Writes}$ trigger a dirty bit update (${\tt W_{db0}}$) corresponding to the PTE of the VA-to-PA mapping.}\label{fig:ghost_write}
    \end{subfigure}
\end{minipage}
\end{minipage}
\caption{ISA ${\tt Read}$ and ${\tt Write}$ instructions invoke additional ghost instructions when executed on systems with VM. The ghost instructions access the PTE stored at address ${\tt z}$ to update the TLB or page table's state.}
\label{fig:ghost}
\end{figure}

User-facing code can cause hardware to execute ghost instructions, such as hardware PT walks, on behalf of a memory access (i.e., ${\tt MemoryEvent}$)~\cite{coatcheck}. Ghost instructions are not fetched, decoded, or issued as part of the program instruction stream. Rather, they are invoked on behalf of a particular user-facing instruction in the pipeline. They interact with user-facing instructions via the shared memory state that they modify, such as PTE status bits and TLB entries.

Since ghost instructions are not fetched and issued like user-facing instructions, they are not related to other ${\tt Events}$ on the same thread by ${\tt po}$.
Instead, we define the 
${\tt ghost}$ relation to relate each user-facing instruction to the ghost instruction(s) invoked on its behalf. 
In the ELT examples in this paper, a ${\tt ghost}$ relation exists between a user-facing instruction and a ghost instruction when both have matching numerical subscripts. For example,
Fig.~\ref{fig:ghost_read} illustrates a single ghost instruction, ${\tt R_{ptw0}}$, invoked by user-facing instruction ${\tt R_0}$. 
We next describe the ghost instructions that \fw currently supports.

\subsubsection{PT walks}
\label{sec:ptw}
${\tt MemoryEvents}$ operate on effective VAs; that is, they are specified to access data at a particular VA by the ${\tt address}$ relation (\S\ref{sec:mcm-vocab}). For each memory access to a particular VA, the processor must use hardware and system support to identify the corresponding PA and physical page. 
If the address mapping needed by a user-facing ${\tt MemoryEvent}$ is not already present in the issuing core's TLB,
a PT walker traverses the system's page tables to locate the mapping in a PTE
and load it into a TLB entry. As with data caches, subsequent accesses to the same mapping can access this mapping from the TLB until it is evicted.
Thus, a PT walk is not required for every memory access, but only those that experience TLB misses. Many systems implement hardware PT walkers for performance, which we assume here. (Our grammar is applicable for software PT walks as well.)

Fig.~\ref{fig:ghost} illustrates how \fw models PT walks and their effects on program behavior. In both subfigures, ${\tt R_{ptw0}}$ is a PT walk that loads the address mapping for VA ${\tt x}$ stored at VA ${\tt z}$. 
PT walks (e.g., ${\tt R_{ptw0}}$) must populate a TLB entry before user-facing
instructions (e.g., ${\tt R_0}$ in Fig.~\ref{fig:ghost_read} and ${\tt W_0}$ in Fig.~\ref{fig:ghost_write}) can use it. 
The  ${\tt rf\_ptw}$ relation is introduced to model this new type of ${\tt rf}$ relationship; it relates a PT walk that loads a mapping into a TLB entry, to all user-facing ${\tt MemoryEvents}$ that ``read from'' that specific TLB entry.
${\tt rf\_ptw}$ differs from ${\tt rf}$ in that the ${\tt Location}$ accessed by the PT walk is an address mapping, whereas the ${\tt Location}$ accessed by the user-facing instruction is a data location.

Each PT walk can only be related to one user-facing instruction with ${\tt ghost}$ (the one that triggered it), but it can be related to several user-facing instructions with ${\tt rf\_ptw}$ (those that use the TLB entry it created). ${\tt MemoryEvents}$ that are (resp. are not) related to a PT walk operation with ${\tt ghost}$ represent TLB misses (resp. hits). As discussed further in \S\ref{sec:invlpg}, the eviction of an address mapping from a TLB will result in a TLB miss for that address mapping in a subsequent memory access. Thus, a ${\tt MemoryEvent}$ that experiences a TLB miss must invoke a PT walk to re-load the required mapping back into the TLB.
Referring back to Fig.~\ref{fig:sb}, each ${\tt MemoryEvent}$ accesses a distinct VA and thus should invoke its own PT walk. The ${\tt sb}$ ELTs in Figs.~\ref{fig:sb-elt-permitted}~
and~\ref{fig:sb-elt-forbidden} feature these PT walks and their relationship via ${\tt rf\_ptw}$ to the user-facing ${\tt MemoryEvents}$ that invoke and ``read from'' them.

\subsubsection{Dirty Bit Updates}
\label{sec:dirty}

When an instruction writes to a memory address, the written data is typically propagated to the cache before it is written back to a physical page in memory. Thus, each PTE contains a dirty bit to indicate when the physical page corresponding to that PTE has been modified and therefore needs to be updated~\cite{intel:x86}.

As with PT walks, ${\tt ghost}$ associates a dirty bit update with the user-facing ${\tt Write}$ that caused it.
Fig.~\ref{fig:ghost_write} shows this relation where 
${\tt W_{db0}}$ is a dirty bit ${\tt Write}$ that accesses the dirty bit in the PTE stored at VA ${\tt z}$. The user-facing ${\tt Write}$ that caused ${\tt W_{db0}}$, namely ${\tt W_0}$, shares the same numerical subscript. Likewise, Figs.~\ref{fig:sb-elt-permitted}~and~\ref{fig:sb-elt-forbidden} both include dirty bit ${\tt Writes}$ related by ${\tt ghost}$ to the highlighted user-facing ${\tt Writes}$.
Dirty bit updates are typically performed as RMW operations~\cite{intel:x86}. However, \fw models dirty bit updates as ${\tt Write}$ operations. This is conservative in terms of event ordering, and it reduces the number of instructions \fw requires to synthesize programs with ${\tt Writes}$ from three (user-facing ${\tt Write}$, dirty bit ${\tt Read}$, dirty bit ${\tt Write}$) to two (user-facing ${\tt Write}$, dirty bit ${\tt Write}$). Furthermore, \fw does not explicitly model the OS updating of dirty bits for synonyms (i.e., VAs that map the same PA as in Fig.~\ref{fig:sb-elt-forbidden}). We assume the OS checks all synonym dirty bits before evicting (i.e., swapping out) pages. This assumption is common in non-naive OSs including Linux and could be relaxed in future implementations of \fw.

\subsection{System-Level Operations: Support Instructions}
\label{sec:remappings}

Support instructions help coordinate software's correct execution on hardware systems implementing VM. In particular, an OS has the ability to modify VA-to-PA mappings or invalidate TLB entries during a user-facing program's execution. These ${\tt Events}$ impact the execution of user-facing program instructions by, for example, influencing which PA is ultimately accessed by a user-facing ${\tt MemoryEvent}$.

User-level MCM relations are traditionally defined assuming that each memory location accessed in a program is either a unique PA or a unique VA with no synonyms. Furthermore, user-level MCM relations do not support VA-to-PA mapping changes during a program's execution. Thus, there is no existing MCM vocabulary for articulating which PA is being accessed by a particular VA and by extension, no vocabulary for modeling synonyms that can arise from system-level ${\tt Events}$. \fw's MTM vocabulary solves both of these issues. 
First, \fw enables expressing the OS's ability to alter VA-to-PA mappings via system calls by introducing support instructions for writing to PTEs as a type of ${\tt Write}$ event. \fw thus supports ${\tt com}$ edges (\S\ref{sec:mcm-vocab}) that
relate ${\tt MemoryEvents}$ with different effective VAs as long as these VAs map to the same PA (i.e., ${\tt com}$ edges relate same-PA ${\tt MemoryEvents}$). Second, \fw accounts for the OS's ability to alter TLB state
by introducing a support instruction for 
evicting specified address mappings from the TLB (${\tt INVLPG}$), either as a result of a system call changing the address mapping or spuriously.

\subsubsection{VA-to-PA Remappings}
\label{sec:pa_relations}

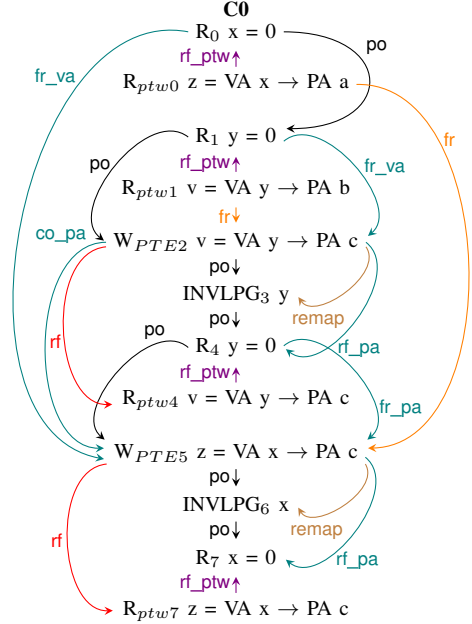
\begin{figure}[t]
\begin{minipage}[b]{\linewidth}
    \begin{subfigure}[b]{\linewidth}
    \centering
	    \footnotesize\setlength{\tabcolsep}{1pt}
	    \begin{tabular}{| l |}
	         \hline
	         \multicolumn{1}{| c |}{\textbf{C0}} \\ \hline
	         R$_{0}$ x = 0 \\
	         \cellcolor{lightgray} \hspace{2mm} R$_{ptw0}$ z = VA x $\rightarrow$ PA a \\
             R$_{1}$ y = 0 \\
             \cellcolor{lightgray} \hspace{2mm} R$_{ptw1}$ v = VA y $\rightarrow$ PA b \\
             W$_{PTE2}$ v = VA y $\rightarrow$ PA c \\
             INVLPG$_{3}$ y \\
             R$_{4}$ y = 0 \\
             \cellcolor{lightgray} \hspace{2mm} R$_{ptw4}$ v = VA y $\rightarrow$ PA c \\
             W$_{PTE5}$ z = VA x $\rightarrow$ PA c \\
             INVLPG$_{6}$ x \\
             R$_{7}$ x = 0 \\
             \cellcolor{lightgray} \hspace{2mm} R$_{ptw7}$ z = VA x $\rightarrow$ PA c \\ \hline
        \end{tabular}
        \caption{ELT with two PTE ${\tt Writes}$ and several ${\tt Reads}$ that read from various mappings, depending on their location in the program. Here and throughout the paper, white cells represent user- or system-level instructions while gray lines represent ghost instructions.}\label{fig:pa_edges_code}
    \end{subfigure}
\end{minipage}
\\
\begin{minipage}[b]{\linewidth}
\begin{subfigure}[b]{\linewidth}
        \centering
        
        \begin{tikzpicture}[->,>=stealth,shorten >=1pt,auto,node distance=.7cm, node/.style={rectangle,draw=none,fill=none,minimum size=1mm}]
        \hspace{-3.5mm}
          \footnotesize
          \node[node] (C0) {\textbf{C0}};
          \node[node] (i0) [below of=C0, yshift=.4cm]    {R$_{0}$ x = 0};
          \node[node] (i1) [below of=i0]   {R$_{ptw0}$ z = VA x $\rightarrow$ PA a};
          \node[node] (i2) [below of=i1]   {R$_{1}$ y = 0};
          \node[node] (i3) [below of=i2]   {R$_{ptw1}$ v = VA y $\rightarrow$ PA b};
          \node[node] (i4) [below of=i3]   {W$_{PTE2}$ v = VA y $\rightarrow$ PA c};
          \node[node] (i5) [below of=i4]   {INVLPG$_{3}$ y};
          \node[node] (i6) [below of=i5]   {R$_{4}$ y = 0};
          \node[node] (i7) [below of=i6]   {R$_{ptw4}$ v = VA y $\rightarrow$ PA c};
          \node[node] (i8) [below of=i7]   {W$_{PTE5}$ z = VA x $\rightarrow$ PA c};
          \node[node] (i9) [below of=i8]   {INVLPG$_{6}$ x};
          \node[node] (i10) [below of=i9]   {R$_{7}$ x = 0};
          \node[node] (i11) [below of=i10]   {R$_{ptw7}$ z = VA x $\rightarrow$ PA c};
        
          \path[every node/.style={font=\sffamily\scriptsize, fill=none,inner sep=2pt}]
            (i0.0) edge [bend left=90, looseness=3, near start, right] node[xshift=5pt] {po} (i2.10)
            (i2) edge [bend right=90, left] node {po} (i4)
            (i4) edge [left] node {po} (i5)
            (i5) edge [left] node {po} (i6)
            (i6.180) edge [bend right=90, near start, above] node {po} (i8.175)
            (i8) edge [left] node {po} (i9)
            (i9) edge [left] node {po} (i10)
            
            (i4.182) edge [red, bend right=88, left] node[yshift=-5pt, xshift=1pt] {rf} (i7.182)
            (i8.185) edge [red, bend right=88, left] node {rf} (i11.180)
            (i1.0) edge [orange, bend left=90, near start, right] node {fr} (i8.2)
            (i3) edge [orange, left] node {fr} (i4)
            (i1) edge [violet, left] node {rf\_ptw} (i0)
            (i3) edge [violet, left] node {rf\_ptw} (i2)
            (i7) edge [violet, left] node {rf\_ptw} (i6)
            (i11) edge [violet, left] node {rf\_ptw} (i10)
            (i4.-2) edge [brown, bend left=90, right, below, near end] node[yshift=-2pt] {remap} (i5.0)
            (i8.-5) edge [brown, bend left=90, right, below, near end] node[yshift=-2pt] {remap} (i9.0)
            (i4) edge [teal, bend left=90, right] node[yshift=-11pt, xshift=-9pt] {rf\_pa} (i6)
            (i8.-2) edge [teal, bend left=90, right] node[yshift=-11pt, xshift=-9pt] {rf\_pa} (i10.0)
            (i6.2) edge [teal, bend left=90, right, near end] node {fr\_pa} (i8.5)
            (i4.180) edge [teal, bend right=90, left, above, near start] node[yshift=10pt] {co\_pa} (i8.178)
            (i0.180) edge [teal, bend right=90, left, near start] node[yshift=2pt] {fr\_va} (i8.183)
            (i2.-5) edge [teal, bend left=90, right] node[xshift=2pt] {fr\_va} (i4.2)
            ;
        \end{tikzpicture}
        \caption{ELT execution corresponding to (\subref{fig:pa_edges_code}), illustrating address mapping changes and resulting ${\tt \_pa}$ edges. }\label{fig:pa_edges_elt}
    \end{subfigure}
\end{minipage}
\caption{Example usage of each of the new ${\tt \_pa}$ edges (\S\ref{sec:pa_relations}). VAs ${\tt x}$ and ${\tt y}$ are accessed before and after their mappings are changed to alias to the same PA.}
\label{fig:pa_edges}
\end{figure}

MTMs support the potential modification 
of VA-to-PA mappings (stored in PTEs) during a program's execution as a result of system calls. The implication of supporting address remappings is that we cannot assume (as MCMs do) that a 
specific VA is mapped to a specific PA throughout the entirety of a program's execution.
Thus, \fw provides new types of communication relations to support reasoning about which PA is accessed by a given user-facing ${\tt MemoryEvent}$. 

To  support modeling and reasoning about the effects of VA-to-PA remappings of user-level program behavior (i.e., to deduce which interactions may take place between ${\tt MemoryEvents}$ with different effective VAs, yet potentially the same effective PA), we adapt the MCM ${\tt com}$ relations from \S\ref{sec:mcm-vocab}. 
\fw's adaptation of ${\tt com}$ edges results in four new relations
that are described as follows. As with program data, \fw represents PAs (and VAs) symbolically.
\begin{itemize}
    \item ${\tt rf\_pa}$: Relates a PTE ${\tt Write}$ of VA ${\tt v}$ $\rightarrow$ PA ${\tt p}$ to ${\tt MemoryEvents}$ that access PA ${\tt p}$ via VA ${\tt v}$.
    \item ${\tt co\_pa}$: Relates PTE ${\tt Writes}$ of VA ${\tt v}$ $\rightarrow$ PA ${\tt p}$ and VA ${\tt v'}$ $\rightarrow$ PA ${\tt p}$ in a total order. In other words, ${\tt co\_pa}$ is a total order on the creation of aliases to a particular PA ${\tt p}$.
    \item ${\tt fr\_pa}$: Relates a ${\tt MemoryEvent}$ that accesses PA ${\tt p}$ via VA ${\tt v}$ to the ${\tt co\_pa}$-successors of the PTE ${\tt Write}$ that it ``reads from'' in ${\tt rf\_pa}$. 
    \item ${\tt fr\_va}$: Relates a ${\tt MemoryEvent}$ that accesses PA ${\tt p}$ via VA ${\tt v}$ to the ${\tt co}$-successors of the PTE ${\tt Write}$ that it ``reads from'' in ${\tt rf\_pa}$.
\end{itemize}
The relations above are used by TransForm to derive ${\tt com}$ edges that relate ${\tt MemoryEvents}$ accessing the same PA but different VAs (\S\ref{sec:mcm-vocab}). Notably, ${\tt rf\_va}$ and ${\tt co\_va}$ are not included in the above list. This is because ${\tt rf\_va}$ and ${\tt co\_va}$ are already captured by ${\tt rf\_pa}$ and ${\tt co}$, respectively.

Fig.~\ref{fig:pa_edges_code} shows a program in which several user-facing ${\tt Reads}$ access either VA ${\tt x}$ or ${\tt y}$. Two PTE ${\tt Write}$ instructions, ${\tt W_{PTE2}}$ and ${\tt W_{PTE5}}$, invoked by system calls remap both ${\tt x}$ and ${\tt y}$ to a new PA ${\tt c}$. ${\tt W_{PTE2}}$ and ${\tt W_{PTE5}}$ then invoke invalidations ${\tt INVLPG_3}$ and ${\tt INVLPG_6}$ (respectively) of the TLB entries corresponding to their remapped VAs to prevent stale mapping accesses. Fig.~\ref{fig:pa_edges_elt} illustrates how these remapping operations relate to each other and to the user-facing program instructions using the ${\tt \_pa}$ edges described above.

System-level PTE ${\tt Writes}$ via system calls can fundamentally change legality of particular MCM litmus test outcomes, as Fig.~\ref{fig:sb-elt-forbidden} shows. Here, ${\tt W_{PTE3}}$ changes the mapping of VA ${\tt y}$ to PA ${\tt a}$ so that VAs ${\tt x}$ and ${\tt y}$ map to the same PA. ${\tt W_{PTE3}}$ is related to ${\tt R_2}$ and ${\tt W_5}$ via ${\tt rf\_pa}$, indicating that ${\tt R_2}$ and ${\tt W_5}$ read from ${\tt W_{PTE3}}$'s new address mapping and thus access the same PA as ${\tt W_0}$ and ${\tt R_6}$. As a result, this particular candidate execution features an illegal coherence violation, as described in \S\ref{sec:mcm_limits}.

\subsubsection{TLB Entry Evictions}
\label{sec:invlpg}

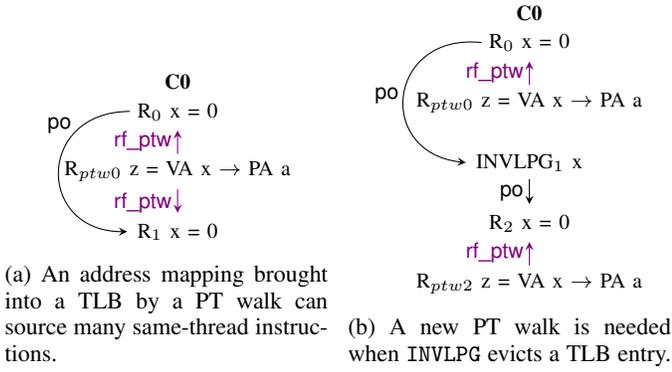
\begin{figure}
    \centering
    \begin{minipage}[b]{\linewidth}
    \begin{minipage}[b]{.485\linewidth}
    \begin{subfigure}[b]{\linewidth}
    \centering
    \begin{tikzpicture}[->,>=stealth,shorten >=1pt,auto,node distance=.8cm, node/.style={rectangle,draw=none,fill=none,minimum size=1mm}]
            \footnotesize
            \node[node] (C0) {\textbf{C0}};
            \node[node] (i0) [below of=C0, yshift=.4cm] {R$_{0}$ x = 0};
            \node[node] (i1) [below of=i0] {R$_{ptw0}$ z = VA x $\rightarrow$ PA a};
            \node[node] (i2) [below of=i1] {R$_{1}$ x = 0};
        
            \path[every node/.style={font=\sffamily\footnotesize, fill=none,inner sep=2pt}]
                (i0) edge [bend right=90, looseness=2, near start, left] node[yshift=2pt] {po} (i2)
                (i1) edge [violet, left] node {rf\_ptw} (i2)
                (i1) edge [violet, left] node {rf\_ptw} (i0)
                ;
        \end{tikzpicture}
    \caption{An address mapping brought into a TLB by a PT walk can source many same-thread instructions.}
    \label{fig:invlpg_a}
    \end{subfigure}
    \end{minipage}
    \hspace{.01\linewidth}
    \begin{minipage}[b]{.485\linewidth}
    \begin{subfigure}[b]{\linewidth}
    \centering
    \begin{tikzpicture}[->,>=stealth,shorten >=1pt,auto,node distance=.8cm, node/.style={rectangle,draw=none,fill=none,minimum size=1mm}]
            \footnotesize
            \node[node] (C0) {\textbf{C0}};
            \node[node] (i0) [below of=C0, yshift=.4cm] {R$_{0}$ x = 0};
            \node[node] (i1) [below of=i0] {R$_{ptw0}$ z = VA x $\rightarrow$ PA a};
            \node[node] (i2) [below of=i1] {INVLPG$_{1}$ x};
            \node[node] (i3) [below of=i2] {R$_{2}$ x = 0};
            \node[node] (i4) [below of=i3] {R$_{ptw2}$ z = VA x $\rightarrow$ PA a};
        
            \path[every node/.style={font=\sffamily\footnotesize, fill=none,inner sep=2pt}]
                (i0) edge [left, bend right=90, looseness=2] node {po} (i2)
                (i1) edge [violet, left] node {rf\_ptw} (i0)
                (i2) edge [left] node {po} (i3)
                (i4) edge [violet, left] node {rf\_ptw} (i3)
                ;
        \end{tikzpicture}
    \caption{A new PT walk is needed when ${\tt INVLPG}$ evicts a TLB entry.}
    \label{fig:invlpg_b}
    \end{subfigure}
    \end{minipage}
    \end{minipage}
    \caption{An ${\tt INVLPG}$ inserted between ${\tt Reads}$ accessing the same VA enforces reloading of the TLB entry via a PT walk.}
    \label{fig:invlpg}
\end{figure}

\S\ref{sec:ptw} discussed how TLB state can be modified via PT Walks. Here, we discuss how TransForm handles TLB state modifications that result from TLB evictions. TLB entries can be evicted for several reasons such as 1) a change in the corresponding PTE, 2) a spurious eviction by the OS, or 3) a TLB capacity eviction. We address each of these scenarios in the following paragraphs.

First, on multicore systems, when address mappings are modified, they must be invalidated in the TLBs of \emph{all} cores caching this mapping, not just the core performing the mapping change. The mechanism used to invoke these page invalidations on each core varies by architecture~\cite{aossvm}. \fw models this TLB entry invalidation using an IPI in the form of an ${\tt INVLPG}$ instruction~\cite{intel:x86}---named for an instruction in the x86 ISA, but similar operations exist in other ISAs as well---related to a PTE ${\tt Write}$ via a ${\tt remap}$ relation. ${\tt remap}$ relates a PTE ${\tt Write}$ to corresponding ${\tt INVLPGs}$ on \emph{each} core that invalidate the TLB entries rendered stale by the PTE ${\tt Write}$ (as in Fig.~\ref{fig:pa_edges}). For example, in Fig.~\ref{fig:sb-elt-forbidden}, ${\tt W_{PTE3}}$ (on ${\tt C1}$) invokes ${\tt INVLPGs}$, ${\tt INVLPG_1}$ and ${\tt INVLPG_4}$, on ${\tt C0}$ and ${\tt C1}$, respectively (as denoted by the ${\tt remap}$ relation), to invalidate the appropriate TLB entries due to the address mapping change. All memory accesses to a VA affected by an ${\tt INVLPG}$ must read from the latest address mapping. In Fig.~\ref{fig:sb-elt-forbidden}, ${\tt R_2}$ and ${\tt W_5}$ read from the new mapping of VA ${\tt y}$ and access PA ${\tt a}$.
Currently, ${\tt INVLPG}$ is the only type of IPI modeled by \fw. However, support for additional IPIs is possible in future \fw extensions.

Second, the OS can initiate TLB evictions even when a PTE has not been modified by a system call (e.g., by spuriously invoking ${\tt INVLPG}$ instructions). 
When an ${\tt INVLPG}$ is invoked by the OS and the corresponding PTE has not changed, TransFrom does not instantiate a ${\tt remap}$ edge. ${\tt MemoryEvents}$ following these spurious ${\tt INVLPGs}$ can read from the unchanged PTE mapping but must bring the mapping back into the TLB with a PT walk.
Fig.~\ref{fig:invlpg_a} illustrates two ${\tt Reads}$, ${\tt R_0}$ and ${\tt R_1}$, to VA ${\tt x}$ that use the mapping brought into the TLB by the same PT walk, ${\tt R_{ptw0}}$. In Fig.~\ref{fig:invlpg_b}, there is an intervening page invalidation, ${\tt INVLPG_1}$, between the two ${\tt Reads}$ to VA ${\tt x}$ so the second ${\tt Read}$, ${\tt R_2}$, invokes a new PT walk, ${\tt R_{ptw2}}$, to bring the previously evicted mapping back into the TLB. When synthesizing ELT candidate executions with \fw's synthesis engine, spurious ${\tt INVLPGs}$ are only inserted on threads if they can affect the thread's execution. 

Finally, TLB capacity evictions occur when a TLB entry must be evicted to make room for a new entry. (This could occur due to capacity or conflict effects.) \fw models these evictions with the invocation of a PT walk by a user-facing ${\tt MemoryEvent}$. As explained in \S\ref{sec:ptw}, the loading of a TLB entry by a PT walk indicates that there was a TLB miss. A TLB miss occurs when 1) the address mapping is first being used, 2) the entry is evicted by the OS (i.e., using ${\tt INVLPG}$), or 3) there is a TLB capacity eviction. Thus, when ELTs feature PT walks that are not accessing an address mapping for the first time (i.e., prior PT walks have been issued for this mapping) and simultaneously do not feature OS evictions of the address mapping from the TLB (i.e., ${\tt INVLPG}$ has not been called for this address mapping), then the PT walk is caused by a TLB capacity eviction. In \fw's automatic synthesis, it explores all three of these possibilities for TLB events.


\subsection{Simplifying Assumptions in ELTs}
Compared to MCM litmus tests, the presence of additional operations, relevant state, and shared memory interactions in ELTs mean that ELTs can become significantly larger (by instruction count) and more complex (by the number of potential interactions between ELT operations). Therefore, the ELTs considered in this paper feature some simplifying assumptions enumerated below. These assumptions do not sacrifice the generality of our approach, but they do result in improved performance when ELTs are analyzed or synthesized using \fw (e.g., in \S\ref{sec:results}).
\begin{enumerate}
    \item Each thread of an ELT is assumed to execute on a distinct processor core. Therefore, each thread has access to private storage, including a private TLB. 
    While 
    \fw can support hyperthreading, the ELTs we consider in this paper represent individual multi-threaded processes as is common practice in MCM analysis.
    \item Prior to the execution of an ELT, it is assumed that each VA maps to a unique PA. Without this assumption, a PTE ${\tt Write}$, along with corresponding ${\tt INVLPGs}$, would need to be explicitly modeled and included in the ELT for \emph{each} VA accessed in the program to appropriately derive which corresponding PAs are accessed. These additional instructions would unnecessarily degrade performance of ELT synthesis with \fw.
    \item We do not model recursive page tables. Stemming from this design choice, we do not model ghost instructions for PTE ${\tt Writes}$. In reality, PTE ${\tt Writes}$ modify some VA whose mapping to a PA is stored in \emph{another} PTE. The process for finding address translations in these higher levels of page tables is very similar to finding them for the base case that we model.
\end{enumerate}

\subsection{Illustrative Example of Vocabulary Usage}

\begin{table}[t]
    \centering
    \begin{tabular}{|c|l|}
        \hline
        \textbf{\begin{tabular}[c]{@{}c@{}}MTM\\ elements\end{tabular}} & \multicolumn{1}{c|}{\textbf{Descriptions}} \\ \hline
        \cellcolor{lightgray}${\tt Event}$ & \cellcolor{lightgray}instruction representing a micro-op in a program \\ \hline
        \cellcolor{lightgray}${\tt MemoryEvent}$ & \cellcolor{lightgray}${\tt Read}$ or ${\tt Write}$ memory access in a program \\ \hline
        \cellcolor{lightgray}${\tt address}$ & \cellcolor{lightgray}relates ${\tt MemoryEvent}$ to ${\tt Location}$ being accessed \\ \hline
        \cellcolor{lightgray}${\tt po}$ & \cellcolor{lightgray}program order, same-thread sequencing of ${\tt Events}$ \\ \hline
        \cellcolor{lightgray}${\tt rf}$ & \cellcolor{lightgray}relates ${\tt Write}$ to ${\tt Reads}$ it sources \\ \hline
        \cellcolor{lightgray}${\tt co}$ & \cellcolor{lightgray}relates ${\tt Write}$ to other ${\tt Writes}$ in coherence order \\ \hline
        \cellcolor{lightgray}${\tt fr}$ & \cellcolor{lightgray}\begin{tabular}[c]{@{}l@{}}relates ${\tt Read}$ to ${\tt co}$-successors of ${\tt Write}$ it reads from\end{tabular} \\ \hline
        ${\tt ghost}$ & \begin{tabular}[c]{@{}l@{}}relates user-facing ${\tt MemoryEvent}$ to induced ghost\\ instructions\end{tabular} \\ \hline
        ${\tt rf\_ptw}$ & \begin{tabular}[c]{@{}l@{}}relates PT walk to user-facing ${\tt MemoryEvents}$ that\\ read from loaded TLB entry\end{tabular} \\ \hline
        ${\tt rf\_pa}$ & \begin{tabular}[c]{@{}l@{}}relates PTE ${\tt Write}$ to user-facing ${\tt MemoryEvents}$\\ that access written address mapping\end{tabular} \\ \hline
        ${\tt co\_pa}$ & \begin{tabular}[c]{@{}l@{}}relates PTE ${\tt Write}$ to other subsequent PTE ${\tt Writes}$ \\ for same PA in coherence order\end{tabular} \\ \hline
        ${\tt fr\_pa}$ & \begin{tabular}[c]{@{}l@{}}relates user-facing ${\tt MemoryEvent}$ to ${\tt co\_pa}$-successors\\ of PTE ${\tt Write}$ they read address mapping from\end{tabular} \\ \hline
        ${\tt fr\_va}$ & \begin{tabular}[c]{@{}l@{}}relates user-facing ${\tt MemoryEvent}$ to subsequent PTE\\ ${\tt Write}$ that changes address mapping for accessed VA\end{tabular} \\ \hline
        ${\tt remap}$ & relates PTE ${\tt Write}$ to invoked ${\tt INVLPGs}$ \\ \hline
    \end{tabular}
    \caption{Summary of MTM vocabulary. The elements in gray are baseline MCM elements that \fw builds on.}
    \label{tab:trans_rels}
\end{table}

\begin{figure}[t]
\begin{minipage}[b]{\linewidth}
\begin{minipage}[b]{\linewidth}
    \begin{subfigure}[b]{\linewidth}
    \centering
	    \footnotesize\setlength{\tabcolsep}{3pt}
	    \begin{tabular}{| l | l |}
	         \hline
	         \multicolumn{1}{| c }{\textbf{C0}}  & \multicolumn{1}{ c |}{\textbf{C1}} \\ \hline
             R$_{0}$ x = 0 & W$_{3}$ x = 1 \\ 
             W$_{PTE1}$ z = VA x $\rightarrow$ PA b & R$_{4}$ x = 1 \\
             W$_{2}$ x = 1 & \\ \hline
        \end{tabular}
        \caption{A user-level test in which a ${\tt Read}$ (${\tt R_0}$) reads 0 at VA ${\tt x}$ and the address mapping for ${\tt x}$ is changed. ${\tt R_4}$ reads a 1 at address ${\tt x}$. }\label{fig:ex_user_code}
    \end{subfigure}
\end{minipage}\\
\begin{minipage}[b]{\linewidth}
    \begin{subfigure}[b]{\linewidth}
    \centering
	    \begin{tikzpicture}[->,>=stealth,shorten >=1pt,auto,node distance=.8cm, node/.style={rectangle,draw=none,fill=none,minimum size=1mm}]
          \footnotesize
          \node[node] (C0) {\textbf{C0}};
          \node[node] (C1) [right of=C0, xshift=3cm] {\textbf{C1}};
          \node[node] (i0) [below of=C0, yshift=.4cm]     {R$_{0}$ x = 0};
          \node[node] (i1) [below of=i0]   {W$_{PTE1}$ z = VA x $\rightarrow$ PA b};
          \node[node] (i2) [below of=i1]   {W$_{2}$ x = 1};
          \node[node] (i3) [below of=C1, yshift=.4cm]   {W$_{3}$ x = 1};
          \node[node] (i4) [below of=i3]   {R$_{4}$ x = 1};

          \path[every node/.style={font=\sffamily\footnotesize, fill=none,inner sep=2pt}]
            (i0) edge [left] node {po} (i1)
            (i1) edge [left] node {po} (i2)
            (i3) edge [right] node {po} (i4)
            (i0) edge [orange] node {fr} (i3)
            (i2) edge [red, below] node {rf?} (i4)
            (i3) edge [red, bend left=90, right] node {rf?} (i4)
            ;
        \end{tikzpicture}
        \caption{Mapping of (\subref{fig:ex_user_code}) showing an ambiguous data access by ${\tt R_4}$.}\label{fig:ex_user}
    \end{subfigure}
\end{minipage}
\end{minipage} \\ \vspace{2pt} \\
\begin{minipage}[b]{\linewidth}
    \begin{subfigure}[b]{\linewidth}
    \centering
	    \footnotesize\setlength{\tabcolsep}{3pt}
	    \begin{tabular}{| l | l |}
	         \hline
	         \multicolumn{1}{| c }{\textbf{C0}}  & \multicolumn{1}{ c |}{\textbf{C1}} \\ \hline
             R$_{0}$ x = 0 & W$_{4}$ x = 1 \\ 
             \cellcolor{lightgray} \hspace{2mm} R$_{ptw0}$ z = VA x $\rightarrow$ PA a & \cellcolor{lightgray} \hspace{2mm} W$_{db4}$ z = VA x $\rightarrow$ PA a \\
             W$_{PTE1}$ z = VA x $\rightarrow$ PA b & \cellcolor{lightgray} \hspace{2mm} R$_{ptw4}$ z = VA x $\rightarrow$ PA a \\
             INVLPG$_{2}$ x & INVLPG$_{5}$ x \\
             W$_{3}$ x = 1 & R$_{6}$ x = 1 \\
             \cellcolor{lightgray} \hspace{2mm} W$_{db3}$ z = VA x $\rightarrow$ PA b & \cellcolor{lightgray} \hspace{2mm} R$_{ptw6}$ z = VA x $\rightarrow$ PA b \\
             \cellcolor{lightgray} \hspace{2mm} R$_{ptw3}$ z = VA x $\rightarrow$ PA b & \\ \hline
        \end{tabular}
        \caption{A possible ELT for the program in (\subref{fig:ex_user_code}).}\label{fig:ex_elt_code}
    \end{subfigure}
\end{minipage}\\
\begin{minipage}[b]{\linewidth}
    \begin{subfigure}[b]{\linewidth}
    \centering
	    \begin{tikzpicture}[->,>=stealth,shorten >=1pt,auto,node distance=.8cm, node/.style={rectangle,draw=none,fill=none,minimum size=1mm}]
          \footnotesize
          \node[node] (C0) {\textbf{C0}};
          \node[node] (C1) [right of=C0, xshift=2.73cm] {\textbf{C1}};
          \node[node] (i0) [below of=C0, yshift=.4cm]    {R$_{0}$ x = 0};
          \node[node] (i1) [below of=i0]   {R$_{ptw0}$ z = VA x $\rightarrow$ PA a};
          \node[node] (i2) [below of=i1]   {W$_{PTE1}$ z = VA x $\rightarrow$ PA b};
          \node[node] (i3) [below of=i2]   {INVLPG$_{2}$ x};
          \node[node] (i4) [below of=i3]   {W$_{3}$ x = 1};
          \node[node] (i5) [below of=i4]   {W$_{db3}$ z = VA x $\rightarrow$ PA b};
          \node[node] (i6) [below of=i5]   {R$_{ptw3}$ z = VA x $\rightarrow$ PA b};
          
          \node[node] (i7) [below of=C1, yshift=.4cm]   {W$_{4}$ x = 1};
          \node[node] (i8) [below of=i7]   {W$_{db4}$ z = VA x $\rightarrow$ PA a};
          \node[node] (i9) [below of=i8]   {R$_{ptw4}$ z = VA x $\rightarrow$ PA a};
          \node[node] (i10) [below of=i9]   {INVLPG$_{5}$ x};
          \node[node] (i11) [below of=i10]   {R$_{6}$ x = 1};
          \node[node] (i12) [below of=i11]   {R$_{ptw6}$ z = VA x $\rightarrow$ PA b};
          
          \path[every node/.style={font=\sffamily\footnotesize, fill=none,inner sep=2pt}]
            (i0.178) edge [bend right=90, looseness=2.2, near end, left] node {po} (i2.178)
            (i2) edge [left] node {po} (i3)
            (i3) edge [left] node {po} (i4)
            
            (i7) edge [bend left=90, looseness=1.2, very near end, right] node[yshift=-5pt] {po} (i10)
            (i10) edge [right] node {po} (i11)
            
            (i5) edge [red, left] node {rf} (i6)
            (i8) edge [red, right] node {rf} (i9)
            (i4) edge [red] node {rf} (i11)
            (i5) edge [red] node {rf} (i12)
            (i1) edge [orange, above] node[xshift=-2pt] {fr} (i8)
            (i0) edge [orange] node {fr} (i7)
            (i9) edge [orange, near start, above] node[xshift=2pt] {fr} (i2)
            (i8.182) edge [blue, very near start, below] node[xshift=5pt] {co} (i2.0)
            (i2.181) edge [blue, bend right=90, left] node {co} (i5.180)
            
            (i1) edge [violet, left] node {rf\_ptw} (i0)
            (i6) edge [violet, bend left=90, left, near start] node {rf\_ptw} (i4)
            (i9.0) edge [violet, bend right=90, looseness=1, right, near end, above] node[yshift=5pt] {rf\_ptw} (i7.2)
            (i12) edge [violet, right] node {rf\_ptw} (i11)
            (i2.184) edge [brown, bend right=90, below, very near end] node[yshift=-2pt] {remap} (i3.180)
            (i2.-2) edge [brown, near end] node {remap} (i10.180)
            
            (i2.182) edge [teal, bend right=90, left, near end] node[xshift=-5pt] {rf\_pa} (i4.182)
            (i2.-5) edge [teal, left] node {rf\_pa} (i11.180)
            (i0.182) edge [teal, bend right=90, left] node[yshift=12pt, xshift=8pt] {fr\_va} (i2.174)
            (i7.185) edge [teal, above, left, bend right=50] node[yshift=7pt] {fr\_va} (i2.5)
            ;
        \end{tikzpicture}
        \caption{ELT illustrating the program in (\subref{fig:ex_elt_code}).}\label{fig:ex_elt}
    \end{subfigure}
\end{minipage}
\caption{The program in (\subref{fig:ex_user_code}) results in an ambiguous MCM relation mapping in (\subref{fig:ex_user}). The ELT in (\subref{fig:ex_elt_code})~and~(\subref{fig:ex_elt}) presents a transistency view of a possible candidate execution.}
\label{fig:ex_conversion}
\end{figure}
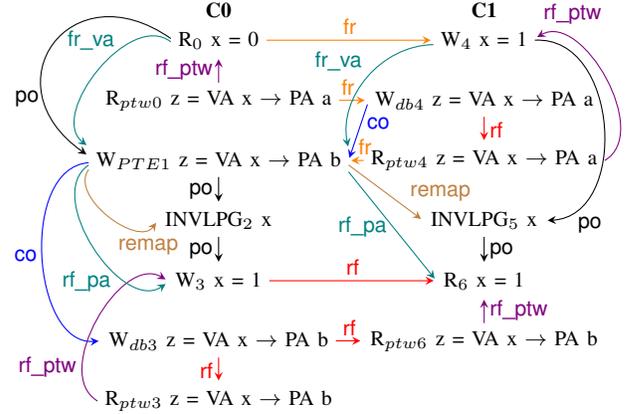

Table~\ref{tab:trans_rels} summarizes \fw's MTM vocabulary. This section illustrates how \fw's MTM vocabulary can be used to reason about observable program behaviors in a more complex ELT example than the running example of Fig.~\ref{fig:motivating}. 

Consider the ELT in Fig.~\ref{fig:ex_conversion}. Fig.~\ref{fig:ex_user_code} features an MCM litmus test
consisting of four user-facing instructions and one support instruction (induced by a system call), ${\tt W_{PTE1}}$.
As Fig.~\ref{fig:ex_user} shows, using only MCM relations results in ambiguity as to which ${\tt Write}$, ${\tt W_2}$ or ${\tt W_3}$, ${\tt R_4}$ is reading from. Although reading from either ${\tt Write}$ constitutes a valid program execution for ${\tt x86\mhyphen{}TSO}$, the ambiguity can be cleared up with MTM relations.

Fig.~\ref{fig:ex_elt_code} extends Fig.~\ref{fig:ex_user_code} to one possible ELT formulation (illustrated with MTM relations in Fig.~\ref{fig:ex_elt}) as follows.
First, Fig.~\ref{fig:ex_user} is augmented with system-level ${\tt INVLPGs}$ that are invoked via OS-issued IPIs on each core. Fig.~\ref{fig:ex_elt} shows ${\tt INVLPG_2}$ and ${\tt INVLPG_5}$ related to ${\tt W_{PTE1}}$ via ${\tt remap}$ relations. Second, ghost instructions are added for each user-facing ${\tt MemoryEvent}$ (${\tt R_0}$, ${\tt W_3}$, ${\tt W_4}$, and ${\tt R_6}$). Since ${\tt INVLPG_2}$ and ${\tt INVLPG_5}$ evict the TLB entries loaded by ${\tt R_{ptw0}}$ and ${\tt R_{ptw4}}$, ${\tt R_{ptw3}}$ and ${\tt R_{ptw6}}$ bring updated address mappings for VA x into the TLB that are accessed by ${\tt W_3}$ and ${\tt R_6}$, respectively. Third, the inserted ghost instructions result in the addition of the appropriate ${\tt \_pa}$ and ${\tt \_va}$ relations (\S\ref{sec:pa_relations}). Since ${\tt W_3}$ and ${\tt R_6}$ access the mapping for VA ${\tt x}$ written by ${\tt W_{PTE1}}$, there are ${\tt rf\_pa}$ edges relating each to ${\tt W_{PTE1}}$. Similarly, ${\tt R_0}$ and ${\tt W_4}$ read from the initial address mapping for VA ${\tt x}$ so there are ${\tt fr\_va}$ edges relating them to ${\tt W_{PTE1}}$. Fourth (and finally), based on the ${\tt \_pa}$ and ${\tt \_va}$ relations, the effective PAs that are accessed by each ${\tt MemoryEvent}$ can be derived and appropriate ${\tt com}$ edges can be added to the ELT. It is now clear that ${\tt R_6}$ reads from ${\tt W_3}$ in this particular execution (due to the ${\tt rf}$ relation between them) while ${\tt W_4}$ accesses a different PA.

As this section's examples  show, Table~\ref{tab:trans_rels}'s proposed MTM vocabulary facilitates modeling of transistency ${\tt Events}$ that encompass consistency issues well beyond traditional MCMs. We next use this vocabulary to automate ELT synthesis.
\section{Automating ELT Synthesis}
\label{sec:elt_synth}

\begin{figure*}
    \centering
    \includegraphics[width=\linewidth]{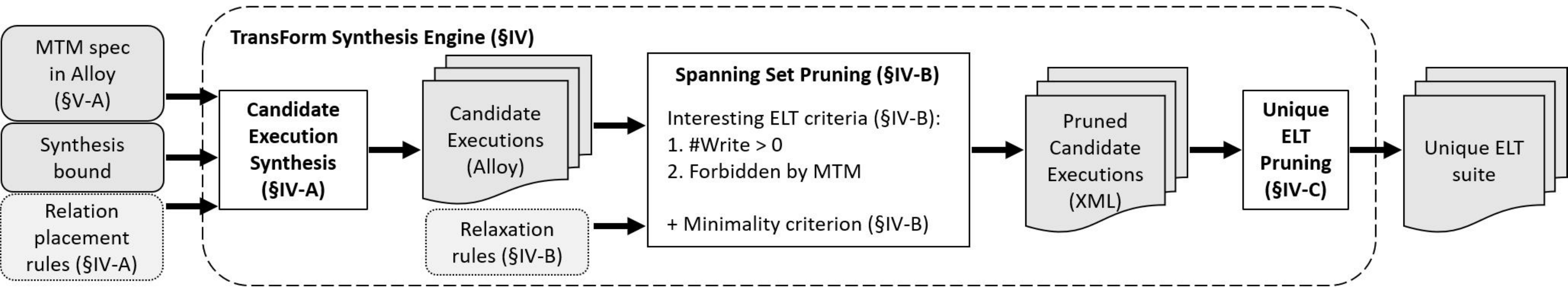}
    \caption{\fw's synthesis engine uses MTM vocabulary from Table~\ref{tab:trans_rels} to axiomatically define the inputted MTM and synthesize candidate executions in Alloy which are pruned and deduplicated to find unique, interesting ELT programs. Relation placement and relaxation rules are implicit inputs that are defined to apply broadly across systems.}
    \label{fig:synth_engine}
\end{figure*}


\fw features a synthesis engine for automatically generating a suite of ELTs from a formal, axiomatic MTM specification supplied using Table~\ref{tab:trans_rels}'s vocabulary. As shown in Fig.~\ref{fig:synth_engine}, \fw's synthesis engine performs bounded ELT synthesis in conceptually three main steps elaborated in the subsections below. First, \fw synthesizes the set of all possible ELT executions up to a user-specified instruction bound. Second, this set of candidate executions is pruned based on which executions feature \textit{interesting} transistency behaviors.
Finally, the subset of interesting candidate executions are deduplicated to output a suite of unique (and interesting) ELT programs. 

\subsection{Candidate Execution Synthesis}
\label{sec:prog_synth}

Synthesizing candidate ELT executions based on our formal MTM vocabulary requires axioms (i.e., rules defined in terms of our vocabulary) that articulate what a valid ELT \textit{looks like}.
Synthesizing traditional MCM litmus tests requires relatively few axioms to describe a legal program execution. For example, consider the MCM features that uniquely define a MCM candidate execution: ${\tt Event}$, ${\tt address}$, ${\tt po}$, ${\tt rf}$, ${\tt co}$, and ${\tt fr}$. One axiom might state that ${\tt po}$ must be acyclic in any valid program execution. Another might state that ${\tt co}$ must represent a total order. To summarize, MCM litmus tests have virtually no constraints on which ${\tt Locations}$ can be related to which ${\tt MemoryEvents}$ via the ${\tt address}$ relation, where individual ${\tt Events}$ can be placed within a program thread, and which (same-address) instructions can interact via ${\tt com}$ relations. As shown in Fig.~\ref{fig:synth_engine}, \fw performs synthesis based on the axioms provided to specify the MTM, as well as a set of placement rules that guide synthesis regarding how operations, such as ghost instructions, can be placed and inserted.   

With \fw's augmented MTM vocabulary  (Table~\ref{tab:trans_rels}), synthesizing candidate ELTs requires
a more complex set of axioms to describe a legal program execution.
For example, ${\tt com}$ edges must relate ${\tt MemoryEvents}$ accessing the \emph{same PA}. PTE ${\tt Writes}$ must induce ${\tt INVLPGs}$ on each core. Likewise, when a program features an ${\tt INVLPG}$, 
a ${\tt MemoryEvent}$ following that ${\tt INVLPG}$ in ${\tt ^{\wedge}po}$ that accesses the address mapping evicted from the local TLB by the ${\tt INVLPG}$ must reload the mapping back into the TLB with a PT walk. Moreover, as described in \S\ref{sec:ghost}, ghost instructions and their corresponding relations (${\tt ghost}$ and ${\tt rf\_ptw}$) have very specific rules that dictate their legal behavior in a candidate execution. 

Given rules for describing legal ELT formulations (as discussed in Section~\ref{sec:towards}), TransForm can synthesize all conceivable ELTs up to a user-specified bound on the number of program instructions. The following sections describe how this set of tests is pruned down to a minimal and interesting subset.


\subsection{Spanning Set Pruning}
\label{sec:spanning_set}

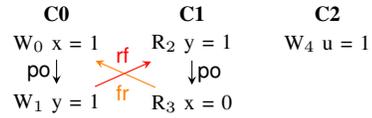
\begin{figure}[t]
\begin{minipage}[b]{\linewidth}
    \begin{subfigure}[b]{\linewidth}
    \centering
    \footnotesize\setlength{\tabcolsep}{3pt}
	    \begin{tabular}{| l | l | l |}
	         \hline
	         \multicolumn{1}{| c }{\textbf{C0}}  & \multicolumn{1}{ c }{\textbf{C1}} &
	         \multicolumn{1}{ c |}{\textbf{C2}}\\ \hline
             W$_{0}$ x = 1 & R$_{2}$ y = 1 & W$_{4}$ u = 1 \\ 
             W$_{1}$ y = 1 & R$_{3}$ x = 0 & \\ \hline
        \end{tabular}
        \caption{A litmus test with ${\tt Writes}$ on ${\tt C0}$ and ${\tt Reads}$ on ${\tt C1}$ to addresses ${\tt x}$ and ${\tt y}$. There is also a ${\tt Write}$ on ${\tt C2}$ to address ${\tt u}$.}\label{fig:nonmin_mp_user_code}
    \end{subfigure}
\end{minipage}
\\
\begin{minipage}[b]{\linewidth}
    \begin{subfigure}[b]{\linewidth}
    \centering
	    \begin{tikzpicture}[->,>=stealth,shorten >=1pt,auto,node distance=.8cm, node/.style={rectangle,draw=none,fill=none,minimum size=1mm}]
	        \footnotesize
              \node[node] (C0) {\textbf{C0}};
              \node[node] (C1) [right of=C0, xshift=1cm] {\textbf{C1}};
              \node[node] (C2) [right of=C1, xshift=1cm] {\textbf{C2}};
            \node[node] (i0) [below of=C0, yshift=.4cm] {W$_{0}$ x = 1};
            \node[node] (i1) [below of=i0] {W$_{1}$ y = 1};
            \node[node] (i2) [below of=C1, yshift=.4cm] {R$_{2}$ y = 1};
            \node[node] (i3) [below of=i2] {R$_{3}$ x = 0};
            \node[node] (i4) [below of=C2, yshift=.4cm] {W$_{4}$ u = 1};
        
            \path[every node/.style={font=\sffamily\footnotesize, fill=none,inner sep=2pt}]
                (i0) edge [left] node {po} (i1)
                (i2) edge [right] node {po} (i3)
                (i1) edge [red] node[xshift=4pt, yshift=2pt] {rf} (i2)
                (i3) edge [orange] node[xshift=4pt, yshift=-2pt] {fr} (i0);
        \end{tikzpicture}
        \caption{Mapping of the program from (\subref{fig:nonmin_mp_user_code}). There is a cycle formed by the ${\tt po}$, ${\tt rf}$, and ${\tt fr}$ edges on ${\tt C0}$ and ${\tt C1}$.}\label{fig:nonmin_mp_user_lt}
    \end{subfigure}
\end{minipage}
\caption{This candidate execution violates ${\tt x86\mhyphen{}TSO}$ axioms described in \S\ref{sec:mcm-vocab}, and therefore would be included in the vector space of interesting ELTs. However, it does not satisfy the minimality criterion; removing the ${\tt Write}$ (${\tt W_4}$) would not make this program satisfiable under ${\tt x86\mhyphen{}TSO}$, even though removing any of the remaining ${\tt Events}$ would. Because it is not minimal, \fw would not synthesize this ELT.}
\label{fig:nonmin_mp}
\end{figure}

\fw's synthesis engine defines and generates a \emph{spanning set} of ELTs. In linear algebra, every vector in a vector space ${\tt V}$ can be written as a linear combination of the vectors in the spanning set ${\tt S}$. In our work, TransForm synthesizes a spanning set ${\tt S}$ of ELTs where the space of all relevant MTM behaviors (that are realizable up to a user-provided instruction bound) can be captured by the ELTs in ${\tt S}$.

\fw requires the following criteria for inclusion of ELTs in the vector space of relevant (i.e., interesting) MTM behaviors. First, an ELT must contain at least one ${\tt Write}$. This requirement enables multiple possible outcomes (i.e., executions) for the ELT. Second, an ELT must be able to produce an outcome that can violate the transistency predicate of the user-provided MTM. This rule ensures that synthesized ELTs have the potential to expose forbidden MTM behaviors when used for verification and validation.

After pruning the set of all legal candidate ELTs to produce only those that belong in our vector space of interesting MTM behaviors, ELTs are evaluated for inclusion in our spanning set based on a \emph{minimality criterion}.
Minimality requires an ELT execution to have a forbidden outcome that becomes legal (according to the transistency predicate) under every possible isolated \textit{relaxation} of the ELT program~\cite{lustig:automated}. For TransForm's synthesis engine, Fig.~\ref{fig:synth_engine} depicts relaxation rules as an implicit input to the synthesis pruning stage. A relaxation corresponds to the removal of an ${\tt Event}$ (or group of ${\tt Events}$ as described below) or dependency\footnote{In our evaluation we only consider ${\tt rmw}$ dependencies, which are modeled as relations that relate the ${\tt Read}$ and ${\tt Write}$ of an RMW operation.} in the ELT. Relaxations are applied to each candidate ELT in the vector space to determine whether it satisfies the minimality criterion. 
Fig.~\ref{fig:nonmin_mp} shows a simplified example of a candidate ELT with only user-facing ${\tt Events}$ presented that \textit{would not} meet the minimality criterion.

The most common relaxation performed by \fw's synthesis engine when evaluating minimality is the removal of an ${\tt Event}$. Conceptually, this relaxation is intended to remove just a \textit{single isolated event}. However, the removal of some ${\tt Events}$ from an ELT may render the ELT invalid.
For example, ghost instructions are not permitted to exist in an ELT if they do not correspond to some user-facing ${\tt MemoryEvent}$ that invokes them. Alternately, some user-facing ${\tt MemoryEvents}$ \emph{require} the invocation of particular ghost instructions. 
Due to these requirements, when performing a relaxation intended to remove a single ${\tt Event}$ TransForm removes additional ${\tt Events}$ to maintain legality of the ELT.
For example, TransForm permits the removal of a ghost instruction if and only if its corresponding  user-facing ${\tt MemoryEvent}$ is itself removed.
Likewise, ${\tt INVLPGs}$ that are invoked 
by a system-level PTE ${\tt Write}$ can only be removed if and only if the PTE ${\tt Write}$ itself is also removed. 
Spurious ${\tt INVLPGs}$ that are \emph{not} a result of PTE changes, however, are free to be removed in isolation.
Unlike the restricted relaxations in this work, user-level MCM litmus test synthesis from prior work permitted relaxations that remove \emph{any} arbitrary ${\tt Event}$ from a litmus test in isolation~\cite{lustig:automated}.



\subsection{Alloy Implementation and Unique ELT Pruning}
\label{sec:unique}
We use the Alloy relational modeling domain-specific language (DSL), specifically Alloy 4.2~\cite{alloy}, to encode axiomatic MTMs written in TransForm's vocabulary and to implement the synthesis engine described in this section.
Alloy's relational model-finding backend, Kodkod~\cite{kodkod}, enables us to transform the ELT synthesis problem into a SAT problem to be fed to any off-the-shelf SAT solver; our experiments use the MiniSat SAT solver~\cite{minisat}. 
Once \fw's synthesis engine determines which ELTs are eligible for inclusion in the spanning set, Alloy outputs them in XML form. XML ELTs are post-processed using a deduplication engine built on prior work to return a set of unique ELT \textit{programs}~\cite{lustig:automated}. Our experiments synthesize ELTs via the process in Fig.~\ref{fig:synth_engine} up to our provided instruction count bounds.

\section{Case Study: x86 MTM}
\label{sec:case_study}

This section uses Table~\ref{tab:trans_rels}'s axiomatic vocabulary to define and develop an MTM, ${\tt x86t\_elt}$, that estimates the MTM of Intel x86 processors, based on a range of public information and analysis from prior work~\cite{intel:x86, coatcheck}. Then, 
\fw's synthesis engine automatically generates the suite of ELTs that encode the spanning set of ${\tt x86t\_elt}$'s MTM behaviors. 

\subsection{Defining x86t\_elt}
\label{sec:x86t_elt}
As with the consistency predicate for ${\tt x86\mhyphen{}TSO}$, the transistency predicate for ${\tt x86t\_elt}$ consists of the conjunction of several axioms. 
Since transistency is a superset of consistency, the axioms that comprise the ${\tt x86t\_elt}$ transistency predicate include, as a subset, the axioms that comprise the ${\tt x86\mhyphen{}TSO}$
consistency predicate (\S\ref{sec:mcm-vocab})~\cite{alglave:herd}. We identify and evaluate two additional ${\tt x86t\_elt}$ transistency axioms, listed below. The first axiom (${\tt \axiom}$) is required for capturing software-visible effects of x86 transistency implementations, while the second ($\tt tlb\_causality$) is a ``diagnostic'' axiom to aid hardware designers in localizing transistency bugs caused by incorrect TLB implementations.
\begin{enumerate}
    \item \textbf{$\tt \axiom$}: The set $\{{\tt fr\_va} + {\tt ^{\wedge}po} + {\tt remap}\}$ of edges must be acyclic.
    \item \textbf{$\tt tlb\_causality$}: The set ${\{\tt ptw\_source + com}\}$ of edges must be acyclic.
\end{enumerate}

The remainder of this section describes the derivation of these  MTM axioms. From analysis of public x86 documentation and prior work~\cite{intel:x86,coatcheck}, we identify forbidden MTM behaviors and use \fw's vocabulary to define axioms that prevent them.

\subsubsection{${\tt \axiom}$}
\label{sec:sc_per_pte}
${\tt \axiom}$ enforces that a ${\tt MemoryEvent}$ ${\tt e}$ must read from the latest VA-to-PA mapping associated with its effective VA when it follows an ${\tt INVLPG}$ ${\tt i}$ in ${\tt ^{\wedge}po}$ and both ${\tt e}$ and ${\tt i}$ access the same VA.
${\tt MemoryEvents}$ can only access PA ${\tt p}$ via VA ${\tt v}$ as long as this 
address mapping remains intact in their local TLB. If a system call remaps VA ${\tt v}$ to some PA ${\tt p'}$ with a PTE ${\tt Write}$ and invokes ${\tt INVLPGs}$ (represented with ${\tt remap}$ relations) on each core, the previous mapping of VA ${\tt v}$ to PA ${\tt p}$ is rendered invalid for ${\tt MemoryEvents}$ following the ${\tt INVLPGs}$ in ${\tt ^{\wedge}po}$.
The relation ${\tt fr\_va}$ is an architecturally-enforced ordering that relates a user-facing ${\tt MemoryEvent}$ to PTE ${\tt Writes}$ that remap its effective VA to a new PA.
${\tt remap}$ represents an architecturally-enforced ordering between a PTE ${\tt Write}$ and the ${\tt INVLPGs}$ it invokes.
Furthermore, 
it is enforced architecturally that a
${\tt MemoryEvent}$ that accesses a TLB entry that was evicted by an ${\tt INVLPG}$ occurring earlier in ${\tt ^{\wedge}po}$ cannot access an ``old'' address mapping. More specifically, ${\tt x86t\_elt}$ enforces that some ${\tt MemoryEvent}$ ${\tt e}$ following some ${\tt INVLPG}$ ${\tt i}$ in ${\tt ^{\wedge}po}$ (where both access the same VA) must access a VA-to-PA mapping that is a ${\tt co}$-successor of the mapping invalidated by ${\tt i}$.
Thus, we require acyclicity of the union of ${\tt fr\_va}$, ${\tt remap}$, and ${\tt ^{\wedge}po}$.

\subsubsection{${\tt tlb\_causality}$}
\label{sec:sc_per_tlb}
${\tt tlb\_causality}$ prevents a causal relationship between some ${\tt MemoryEvent}$ ${\tt e'}$ and some other ${\tt MemoryEvent}$ ${\tt e}$ whose corresponding PT walk sourced the TLB entry accessed by ${\tt e'}$.
Since ${\tt MemoryEvents}$ locate VA-to-PA mappings in the TLB of their local core, ${\tt rf\_ptw}$ represents an architecturally-enforced ordering between a ${\tt MemoryEvent}$ and the PT walk that populates the TLB entry it accesses. 
Furthermore, our ${\tt x86t\_elt}$ model assumes
an architecturally-enforced ordering between the user-facing ${\tt MemoryEvent}$ that invokes (i.e., is related by ${\tt ghost}$ to) a PT walk and other user-facing ${\tt MemoryEvents}$ (on the same core) that access the TLB entry populated by this PT walk.
To represent this ordering relationship, we derive ${\tt ptw\_source}$ to relate a user-facing ${\tt MemoryEvent}$ that invokes a PT walk 
to all other user-facing ${\tt MemoryEvents}$ that are related to this PT walk by ${\tt rf\_ptw}$.
Thus, some ${\tt MemoryEvent}$ ${\tt e'}$ that is ordered after some other ${\tt MemoryEvent}$ ${\tt e}$ in ${\tt ptw\_source}$ cannot be related to ${\tt MemoryEvent}$ ${\tt e}$ by a causal communication relationship. 

As noted above, we include ${\tt tlb\_causality}$ in ${\tt x86t\_elt}$ for the purpose of diagnosing hardware bugs in TLB implementations. In particular, the architecturally-visible effects of ${\tt tlb\_causality}$ violations are already subsumed by violations of another ${\tt x86t\_elt}$ axiom, specifically ${\tt causality}$ (hence the naming convention). However, including ${\tt tlb\_causality}$ enables TransForm to specifically identify which ELTs may be used by hardware verification engineers to localize transistency bugs to incorrectly implemented TLBs.
Of the 140 unique ELTs that \fw synthesizes for ${\tt x86t\_elt}$ (\S\ref{sec:results}), five can be attributed to violations of ${\tt tlb\_causality}$. 

\subsection{Synthesis Approach}
\label{sec:approach}
Given ${\tt x86t\_elt}$ as defined in \S\ref{sec:x86t_elt} as input, \fw's synthesis engine generates a suite of ELT programs.
The synthesized ELTs must constitute a forbidden program execution (according to the ${\tt x86t\_elt}$ transistency predicate in this case) 
that becomes permitted under every possible relaxation.
To synthesize ELTs that can result in forbidden outcomes, we identify (in turn) each of the axioms that comprise ${\tt x86t\_elt}$ (i.e., ${\tt sc\_per\_loc}$, ${\tt rmw\_atomicity}$, ${\tt causality}$, ${\tt \axiom}$, ${\tt tlb\_causality}$) as an axiom to be violated (and thus render the synthesized ELT executions forbidden). Synthesizing an ELT that violates one of these axioms directly corresponds to synthesizing a forbidden ELT execution.

We synthesize five ELT suites, each containing tests that correspond to violations of one of the five ${\tt x86t\_elt}$ axioms, 
for increasing instruction bounds under a one week timeout period. Each suite requires a minimum instruction bound of 4 instructions or higher, depending on the number of instructions needed to form interactions that can violate the respective axiom and constitute the test a part of the spanning set. Thus, synthesis begins at 4 instructions and increases until timeout.
\section[title]{Results\footnotemark}
\label{sec:results}
\footnotetext{This is an updated version of the \fw paper that features updated results reflecting performance optimizations and software bug fixes.}

We supplied our ${\tt x86t\_elt}$ MTM (from \S\ref{sec:case_study}) consisting of five high-level axioms to \fw's synthesis engine. We evaluated TransForm's ability to synthesize spanning sets of ELTs for ${\tt x86t\_elt}$ with increasing instruction counts for a one week timeout period. For each axiom, we found the minimum required instruction bound that results in synthesized ELTs---between four and seven instructions for the axioms shown here---and incrementally increased this bound until synthesis did not terminate in one week's time. 
The resulting synthesized ELTs for each axiom are collected into a set referred to as a ``per-axiom suite.''
The rest of this section details  our synthesis observations, compares the \fw-generated ELTs to a baseline (the hand-generated COATCheck ELT suite~\cite{coatcheck}), and gives examples of synthesized ELTs.

\subsection{Overview of Synthesized Suite}
\label{sec:results_overview}

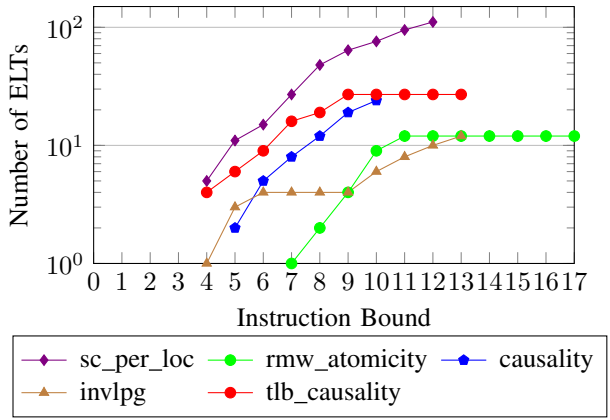
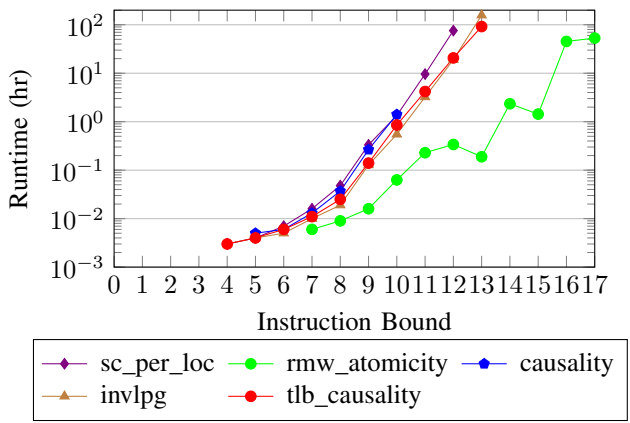
\begin{figure}[t]
\begin{minipage}[b]{\linewidth}
    \begin{subfigure}{\linewidth}
    \centering
	    \begin{tikzpicture}
	    \begin{scope}
        \begin{semilogyaxis}[
            xlabel={Instruction Bound},
            ylabel={Number of ELTs},
            xmin=0, xmax=17,
            ymin=1, ymax=150,
            xtick={0, 1, 2, 3, 4, 5, 6, 7, 8, 9, 10, 11, 12, 13, 14, 15, 16, 17},
            ytickten={0, 1, 2, 3},
            ymajorgrids=true,
            legend columns = 3,
            legend style = {at={(0.45, -0.28)}, anchor=north, inner sep=3pt, style={column sep=0.1cm}},
            legend cell align=left,
            width=.9\linewidth,
            height=5cm
        ]
            
        \addplot[
            color=violet,
            mark=diamond*,
            ]
            coordinates {
            (4,5)(5,11)(6,15)(7,27)(8,48)(9,64)(10,76)(11,95)(12,111)
            };
            \addlegendentry{sc\_per\_loc};
            
        \addplot[
            color=green,
            mark=oplus*,
            ]
            coordinates {
            (7,1)(8,2)(9,4)(10,9)(11,12)(12,12)(13,12)(14,12)(15,12)(16,12)(17,12)
            };
            \addlegendentry{rmw\_atomicity};
            
        \addplot[
            color=blue,
            mark=pentagon*,
            ]
            coordinates {
            (5,2)(6,5)(7,8)(8,12)(9,19)(10,24)
            };
            \addlegendentry{causality};
            
        
        \addplot[
            color=brown,
            mark=triangle*,
            ]
            coordinates {
            (4,1)(5,3)(6,4)(7,4)(8,4)(9,4)(10,6)(11,8)(12,10)(13,12)
            };
            \addlegendentry{\makebox[0pt][l]{\axiom}};
        
        \addplot[
            color=red,
            mark=*,
            ]
            coordinates {
            (4,4)(5,6)(6,9)(7,16)(8,19)(9,27)(10,27)(11,27)(12,27)(13,27)
            };
            \addlegendentry{\makebox[0pt][l]{tlb\_causality}};
            
            
        
        ]
            
        \end{semilogyaxis}
        \end{scope}
        \end{tikzpicture}

        \caption{Plot of the number of ELTs synthesized in each suite by instruction bound. The first point for each type of suite corresponds to the minimum instructions required to synthesize that type of suite.}
        \label{fig:elt_count}
    \end{subfigure}
\end{minipage} \\ \\ \\
\begin{minipage}[b]{\linewidth}
    \begin{subfigure}{\linewidth}
    \centering
	    \begin{tikzpicture}
	    \begin{scope}
        \begin{semilogyaxis}[
            xlabel={Instruction Bound},
            ylabel={Runtime (hr)},
            xmin=0, xmax=17,
            ymin=.001, ymax=200,
            xtick={0, 1, 2, 3, 4, 5, 6, 7, 8, 9, 10, 11, 12, 13, 14, 15, 16, 17},
            ytickten={-3, -2, -1, 0, 1, 2, 3},
            ymajorgrids=true,
            legend columns = 3,
            legend style = {at={(0.45, -0.28)}, anchor=north, inner sep=3pt, style={column sep=0.1cm}},
            legend cell align=left,
            width=.9\linewidth,
            height=5cm
        ]
         
        \addplot[
            color=violet,
            mark=diamond*,
            ]
            coordinates {
            (4,.003)(5,.004)(6,.007)(7,.016)(8,.048)(9,.333)(10,1.343)(11,9.595)(12,75.722)
            };
            \addlegendentry{sc\_per\_loc};
            
        \addplot[
            color=green,
            mark=oplus*,
            ]
            coordinates {
            (7,.006)(8,.009)(9,.016)(10,.063)(11,.229)(12,.339)(13,.189)(14,2.344)(15,1.434)(16,45.325)(17,53.017)
            };
            \addlegendentry{rmw\_atomicity};
            
        \addplot[
            color=blue,
            mark=pentagon*,
            ]
            coordinates {
            (5,.005)(6,.006)(7,.013)(8,.037)(9,.265)(10,1.410)
            };
            \addlegendentry{causality};
            
            
        \addplot[
            color=brown,
            mark=triangle*,
            ]
            coordinates {
            (4,.003)(5,.004)(6,.005)(7,.010)(8,.019)(9,.128)(10,.549)(11,3.231)(12,19.047)(13,157.420)
            };
            \addlegendentry{\makebox[0pt][l]{\axiom}};
            
            
        \addplot[
            color=red,
            mark=*,
            ]
            coordinates {
            (4,.003)(5,.004)(6,.006)(7,.011)(8,.025)(9,.140)(10,.852)(11,4.192)(12,20.681)(13,91.812)
            };
            \addlegendentry{\makebox[0pt][l]{tlb\_causality}};
            
            
            
        
            
        \end{semilogyaxis}
        \end{scope}
        \end{tikzpicture}

        \caption{Plot of the runtimes for synthesizing each suite by instruction bound. Runtimes increase monotonically with instruction bound, with the exception of the ${\tt rmw\_atomicity}$ suite which resulted in non-monotonic runtime behavior when bounds were increased past 11 instructions. For these runs, the same ELTs were synthesized so this non-monotonicity likely occurred due to the inherent variability and noise present during synthesis. Although runtimes grow super-exponentially with instruction bound, our ELT optimizations (\S\ref{sec:towards}) and symmetry reduction enable synthesis for 10-instruction bounds and higher to result in over 100 useful ELTs within practical runtimes. We believe future work on symmetry reductions and other optimizations can further accelerate these synthesis times.}
        \label{fig:elt_runtime}
    \end{subfigure}
\end{minipage} \\
\caption{Statistics on our synthesized  ELTs. (\subref{fig:elt_count}) plots the number of instructions in each synthesized test suite while~(\subref{fig:elt_runtime}) plots the runtimes for synthesizing each of them.}
\label{fig:elt_plots}
\end{figure}

Fig.~\ref{fig:elt_count} summarizes the number of ELT programs synthesized in each per-axiom suite, at varying instruction bounds. Fig.~\ref{fig:elt_runtime} shows the corresponding execution time required to synthesize them.
Instruction bounds with no data points were either too restrictive to synthesize ELTs satisfying our spanning set criteria (i.e., fewer than 4 instructions) or too permissive (resulting in a large search space) for synthesis to complete within one week. Variability among our evaluated axioms results in variable minimal instruction bounds to produce non-empty ELT spanning sets and variable synthesis runtimes.

Over one hundred ELTs are generated automatically. At each instruction bound, the ${\tt sc\_per\_loc}$ suite makes up the largest component of the full synthesized suite. This is in part because the ${\tt sc\_per\_loc}$ axiom specifies ordering constraints on \emph{all} types of instructions that \fw models (i.e., user-facing, support, and ghost). 
This translates into more possibilities for violating this axiom, therefore more ELTs that qualify as interesting based on our criteria for vector space inclusion (in \S\ref{sec:spanning_set}).  
Prior automated MCM litmus test synthesis~\cite{lustig:automated} resulted in the ${\tt sc\_per\_loc}$ suite saturating at 10 tests for the ${\tt x86\mhyphen{}TSO}$ MCM. Because of the richer interactions in MTMs, many more tests are generated in our corresponding synthesis runs here for ${\tt x86t\_elt}$. Overall, the value of \fw's ELT synthesis is two-fold.  First, the automatic synthesis of hundreds of minimal and interesting MTM ELTs offers huge support for systems programmers and transistency verifiers.  Second, the methodology rests on a foundational definition of {\em completeness} up to the specified synthesis bounds; this gives designers a clear understanding of the comprehensiveness of their verification approach.

The \fw synthesis approach generates a complete suite of ELTs, composed of per-axiom suites at varying synthesis bounds, for the provided one week timeout bound.

\subsection{Comparison Against Prior Work}
\label{sec:comparison}

For reference, we compare our \fw-synthesized ELT suite for ${\tt x86t\_elt}$ with a handwritten suite of 40 ELTs from prior work~\cite{coatcheck}. Of the original 40, 22 ELTs are relevant for comparison. Of the other 18, 9 deal with particular IPIs that are not presently supported by \fw, and 9 others do not meet \fw's spanning set criteria for ELTs. In contrast, \fw synthesizes a total of 140 unique ELTs across all per-axiom suites (for a one week synthesis timeout bound).

To facilitate comparison, we consider the 22 prior handwritten ELTs as two categories. First, there are ELTs which pass the minimality criterion and would be synthesized verbatim by \fw (category 1). Second, there are ELTs which are not minimal as-is but are a superset of a minimal ELT (category 2). The extraneous instructions in the latter set of ELTs can be removed, exposing a minimal ELT that \fw would synthesize. We automate the ELT comparison process via a tool that first checks if \fw would synthesize the ELT verbatim in the synthesized suite (category 1), and if not, subsequently tests for the ELT's inclusion in category 2 by trying to remove subsets of instructions from the ELT to see if it can be minimized to a \fw-synthesizable test.

Seven of the 22 ELTs from the COATCheck suite fall into the first category and are synthesized verbatim. These seven ELTs match four synthesized ELT programs. Recall that our tool outputs ELT \emph{programs} whereas ELTs typically describe programs \emph{and} their \emph{outcomes} (i.e., an ELT execution), so some of our synthesized ELT programs might correspond to more than one ELT execution from 
the COATCheck suite. 
We additionally find 15 ELTs from the COATCheck suite that fall into the second category. These ELTs can be reduced to at least 1 minimal ELT which is synthesized by \fw. We consider such synthesized minimal ELTs to be unique new ELTs, as they were not explicitly part of the handwritten COATCheck suite.
Thus, of the 140 ELTs synthesized by \fw, we find that all 22 ELTs from the COATCheck suite either directly match 4 of the synthesized ELTs or derive from one of the remaining 136 new synthesized ELTs.

\subsection{Examples of Synthesized ELTs}
\label{sec:example_results}

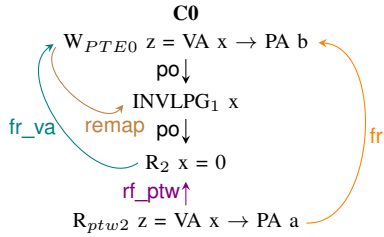
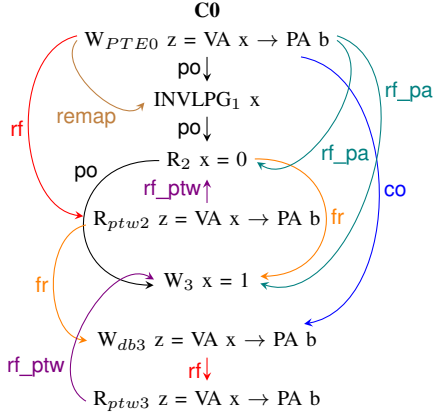
\begin{figure}[t]
\begin{minipage}[b]{\linewidth}
\begin{minipage}[b]{\linewidth}
    \begin{subfigure}[b]{\linewidth}
    \centering
	    \begin{tikzpicture}[->,>=stealth,shorten >=1pt,auto,node distance=.8cm, node/.style={rectangle,draw=none,fill=none,minimum size=1mm}]
          \footnotesize
          \node[node] (C0) {\textbf{C0}};
          \node[node] (i0) [below of=C0, yshift=.4cm]    {W$_{PTE0}$ z = VA x $\rightarrow$ PA b};
          \node[node] (i1) [below of=i0]   {INVLPG$_{1}$ x};
          \node[node] (i2) [below of=i1]   {R$_{2}$ x = 0};
          \node[node] (i3) [below of=i2]   {R$_{ptw2}$ z = VA x $\rightarrow$ PA a};
          
          \path[every node/.style={font=\sffamily\footnotesize, fill=none,inner sep=2pt}]
            (i0) edge [left] node {po} (i1)
            (i1) edge [left] node {po} (i2)
            
            (i3) edge [orange, bend right=90, right] node {fr} (i0)
            
            (i3) edge [violet, left] node {rf\_ptw} (i2)
            (i0.182) edge [brown, bend right=90, left, below, very near end] node[yshift=-2pt] {remap} (i1.180)
            
            (i2) edge [teal, bend left=90, left] node {fr\_va} (i0)
            ;
        \end{tikzpicture}
        \caption{This figure illustrates the forbidden ${\tt ptwalk2}$ ELT from the COATCheck suite that \fw synthesized.}\label{fig:res_ptwalk2}
    \end{subfigure}
\end{minipage}
\\
\begin{minipage}[b]{\linewidth}
    \begin{subfigure}[b]{\linewidth}
    \centering
	    \begin{tikzpicture}[->,>=stealth,shorten >=1pt,auto,node distance=.8cm, node/.style={rectangle,draw=none,fill=none,minimum size=1mm}]
          \footnotesize
          \node[node] (C0) {\textbf{C0}};
          \node[node] (i0) [below of=C0, yshift=.4cm]    {W$_{PTE0}$ z = VA x $\rightarrow$ PA b};
          \node[node] (i1) [below of=i0]   {INVLPG$_{1}$ x};
          \node[node] (i2) [below of=i1]   {R$_{2}$ x = 0};
          \node[node] (i3) [below of=i2]   {R$_{ptw2}$ z = VA x $\rightarrow$ PA b};
          \node[node] (i4) [below of=i3]   {W$_{3}$ x = 1};
          \node[node] (i5) [below of=i4]   {W$_{db3}$ z = VA x $\rightarrow$ PA b};
          \node[node] (i6) [below of=i5]   {R$_{ptw3}$ z = VA x $\rightarrow$ PA b};
          
          \path[every node/.style={font=\sffamily\footnotesize, fill=none,inner sep=2pt}]
            (i0) edge [left] node {po} (i1)
            (i1) edge [left] node {po} (i2)
            (i2.180) edge [bend right=90, looseness=2, left, near start] node[yshift=2pt] {po} (i4.185)
            
            (i2.2) edge [orange, bend left=90, looseness=2] node {fr} (i4.5)
            (i3.182) edge [orange, bend right=80, left] node {fr} (i5.180)
            (i5) edge [red, left] node {rf} (i6)
            (i0) edge [blue, bend left=80] node {co} (i5)
            (i0.178) edge [red, bend right=80, left] node {rf} (i3.178)
            
            (i3) edge [violet, left] node {rf\_ptw} (i2)
            (i6) edge [violet, bend left=90, near start] node {rf\_ptw} (i4)
            (i0) edge [brown, bend right=90, left, near end] node[yshift=-5pt] {remap} (i1)
            
            (i0) edge [teal, bend left=90, right, near end] node[xshift=8pt, yshift=2pt] {rf\_pa} (i2)
            (i0.2) edge [teal, bend left=90, right, near start] node {rf\_pa} (i4.-2)
            ;
        \end{tikzpicture}
        \caption{This figure illustrates the permitted ${\tt dirtybit3}$ ELT from the COATCheck suite that can be reduced to a program that meets the minimality criterion and can be synthesized by \fw.}\label{fig:res_dirtybit3}
    \end{subfigure}
\end{minipage}
\end{minipage}
\caption{(\subref{fig:res_ptwalk2}) and (\subref{fig:res_dirtybit3}) illustrate COATCheck ELTs that were synthesized by \fw either verbatim or by removing extraneous instructions, respectively.}
\label{fig:res_elts}
\end{figure}

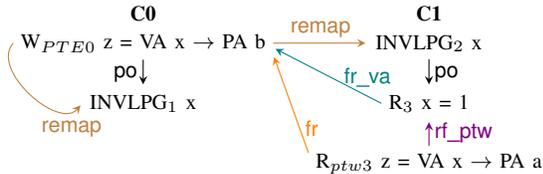
\begin{figure}[t]
    \centering
    \begin{tikzpicture}[->,>=stealth,shorten >=1pt,auto,node distance=.8cm, node/.style={rectangle,draw=none,fill=none,minimum size=1mm}]
      \footnotesize
      \node[node] (C0) {\textbf{C0}};
      \node[node] (C1) [right of=C0, xshift=3cm] {\textbf{C1}};
      \node[node] (i0) [below of=C0, yshift=.4cm]     {W$_{PTE0}$ z = VA x $\rightarrow$ PA b};
      \node[node] (i1) [below of=i0]   {INVLPG$_{1}$ x};

      \node[node] (i4) [below of=C1, yshift=.4cm]   {INVLPG$_{2}$ x};
      \node[node] (i5) [below of=i4]   {R$_{3}$ x = 1};
      \node[node] (i7) [below of=i5]   {R$_{ptw3}$ z = VA x $\rightarrow$ PA a};
      
      \path[every node/.style={font=\sffamily\footnotesize, fill=none,inner sep=2pt}]
        (i0) edge [left] node {po} (i1)
        
        (i4) edge [right] node {po} (i5)
        
        (i5.180) edge [teal, near start, above] node[xshift=5pt] {fr\_va} (i0.-2)
        (i7.175) edge [orange, near start, right] node {fr} (i0.-5)
        
        (i0) edge [brown, bend right=90, below, very near end] node[yshift=-2pt] {remap} (i1)
        (i0.0) edge [brown, above] node {remap} (i4)
        
        (i7) edge [violet, right] node {rf\_ptw} (i5)
        ;
    \end{tikzpicture}
    \caption{This figure illustrates a forbidden candidate execution of a new synthesized ELT.}
    \label{fig:res_new}
\end{figure}

Figs.~\ref{fig:res_ptwalk2} and~\ref{fig:res_new} provide examples of synthesized ELTs. Fig.~\ref{fig:res_ptwalk2} illustrates an ELT synthesized by \fw and an exact match to a category 1 example (${\tt ptwalk2}$) from the COATCheck suite. The outcome shown violates both ${\tt sc\_per\_loc}$ and ${\tt \axiom}$ so it is forbidden. 

Fig.~\ref{fig:res_dirtybit3} illustrates a handwritten ELT from the COATCheck suite. It is one of the 15 category 2 ELTs in our comparison, meaning it is not minimal in its handwritten form. \fw automatically synthesizes a reduction of this ELT. Our comparison tool identifies a possible reduction (i.e., a subset of extraneous instructions that can be removed, in this case ${\tt \{W_3}\}$) of this ELT
that renders it minimal and synthesizable. 


Fig.~\ref{fig:res_new} illustrates an example of a new synthesized ELT that is not found in the handwritten suite. In it, a system-level PTE ${\tt Write}$, ${\tt W_{PTE0}}$, is invoked by a system call, and remaps VA ${\tt x}$.
${\tt W_{PTE0}}$'s mapping update induces two ${\tt INVLPGs}$: ${\tt INVLPG_1}$ and ${\tt INVLPG_2}$.
${\tt INVLPG_2}$ precedes ${\tt R_3}$ which reads from VA ${\tt x}$. This particular execution has a forbidden outcome because even though ${\tt R_3}$ comes after ${\tt INVLPG_2}$ in ${\tt po}$, it accesses a stale address mapping, as indicated by ${\tt fr\_va}$. This execution violates ${\tt \axiom}$ since there is a resulting cycle in ${\tt remap}$, ${\tt fr\_va}$, and ${\tt ^{\wedge}po}$, and is thus forbidden by ${\tt x86t\_elt}$.

\fw's framework for specifying MTMs and automatically synthesizing ELTs paves the way for systems programmers to perform deep verification and validation of MTM implementations. 
In future work, we plan to use the synthesized ELTs to empirically validate ${\tt x86t\_elt}$  against real-world operating systems and x86 processor implementations.
\section{Related Work}
\label{sec:related}
\noindent\textbf{Formal MCM Specifications:} From their earliest roots \cite{lamport:sc}, MCMs have been studied extensively over the years. Programming language-level MCMs have been formalized for Java, C11, and OpenCL~\cite{manson:java, boehm:cppconcurrency, petri:cooking, Batty:mathematizingc++, batty:overhauling, nienhuis:c11operational, Wickerson2015}. Additionally, formal ISA-level MCM specifications exist for x86-TSO, Power, ARMv7, ARMv8, RISC-V WMO and TSO, and NVIDIA PTX~\cite{nagarajan2020primer, owens:better, alglave:herd, pulte:armv8, nvidia:ptx, RISCV:rvtso:rvwmo}. These MCM specification efforts have given way to verified compiler mapping schemes from C11 and Java MCM primitives to the x86, ARMv7, ARMv8, and Power ISAs~\cite{Batty:mathematizingc++, batty:c++topower, POWERlrsc, lahav:repairing, sewell:mappings, petri:cooking, sevcik:javacompiler}. 
Recently, the MCM for the Linux Kernel was also formalized~\cite{alglave:linux}. \fw assists programmers and compiler writers in developing correct system code for VM implementations by offering specification and verification support.

\noindent\textbf{Verification of Hardware MCM Implementations:}
Formal ISA MCM specifications have prompted research on verifying the correctness of hardware MCM implementations~\cite{pipecheck,ccicheck,tricheck,rtlcheck,pipeproof}. Much of this prior work conducts bounded verification for suites of MCM litmus tests~\cite{pipecheck,ccicheck,tricheck,rtlcheck} while some is proof-based~\cite{pipeproof,choi:kami}. The COATCheck tool from this line of work also proposed a mechanism for verifying MTM implementations~\cite{coatcheck}. However, this work relied on hand-crafted ELTs and did not formally describe them or an MTM which could be used to generate them. In contrast, \fw can be used to generate ELTs for expanding coverage of hardware MTM verification.

\noindent\textbf{Expanding the Scope of Concurrency Specifications:}
A variety of research efforts formally define concurrency specifications beyond memory consistency. \textit{Crash consistency} has been proposed to describe ordering behaviors for file system state updates across crashes~\cite{bornholt:crash}. \textit{Memory persistency} has been coined for reasoning about the order in which nonvolatile memory writes persist to memory~\cite{pelley:persistency}. Recently, persistency models have been formalized in the context of the release consistency~\cite{izraelevitz:prelease}, x86-TSO~\cite{raad:ptso}, and ARMv8~\cite{raad:parm} MCMs. 
\section{Conclusion}
\label{sec:conclusion}

\fw is a framework for formally specifying MTMs and synthesizing ELTs to support systems programmers and hardware designers verifying MTM behaviors. MTMs are central for assuring correct consistency behavior in the face of intricate VM interactions. \fw includes a vocabulary for formally specifying MTMs and an automated synthesis approach for corresponding ELTs. To evaluate \fw, we used its vocabulary to specify ${\tt x86t\_elt}$, an estimated MTM for x86 processors.  From this MTM, we used \fw to automatically synthesize its corresponding ELTs.  TransForm’s synthesis engine automatically produces a
 set of ELTs including relevant
hand-curated ELTs from prior work, plus over 100 more. \fw is open source and publicly available at github.com/naorinh/TransForm.git.
Overall, this work showcases the value and potential impact \fw brings to MTM verification. 

\section*{Acknowledgments}

Thanks to Daniel Lustig and Yatin Manerkar for helpful feedback. This work was supported in part by Intel Corp. and through NSF CISE XPS-16-28926.


\bibliographystyle{IEEEtranS}
\bibliography{refs}

\end{document}